\documentclass[twocolumn,prd,aps,superscriptaddress,preprintnumbers,tightenlines,showpacs,nofootinbib,eqsecnum,amsfonts,amsmath]{revtex4-1}
\usepackage{placeins}
\usepackage{color}
\usepackage{calc}
\usepackage{amsmath,amssymb,graphicx}
\usepackage{tensor}
\usepackage{bm}
\usepackage{times}
\usepackage[varg]{txfonts}
\usepackage[colorlinks, pdfborder={0 0 0}]{hyperref}
\usepackage{float}
\usepackage{dcolumn}
\usepackage[nolist,nohyperlinks]{acronym}
\usepackage{xspace}
\usepackage[abs]{overpic}
\usepackage{pict2e}
\usepackage{enumitem}
\usepackage[usenames,dvipsnames]{xcolor}
\usepackage[utf8]{inputenc}
\usepackage{acronym}
\usepackage{gensymb}
\usepackage{cleveref}
\usepackage[normalem]{ulem}
\usepackage{longtable}
\usepackage{multirow}

{
\newcommand{\chirpm}{\mathcal{M}_c}
\usepackage{subfigure}

\usepackage{placeins}

\usepackage{booktabs}

\newcommand{\bsub}{\begin{subequations}}
\newcommand{\esub}{\end{subequations}}

\begin{document}

 \title{Parameter estimation of stellar-mass black hole binaries with LISA}

\author{Alexandre Toubiana}
\affiliation{APC, AstroParticule et Cosmologie,
Université de Paris, CNRS, Astroparticule et Cosmologie, F-75013 Paris, France}
\affiliation{Institut d'Astrophysique de Paris, CNRS \& Sorbonne
 Universit\'es, UMR 7095, 98 bis bd Arago, 75014 Paris, France}

\author{Sylvain Marsat}
\affiliation{APC, AstroParticule et Cosmologie,
Université de Paris, CNRS, Astroparticule et Cosmologie, F-75013 Paris, France}

\author{Stanislav Babak}
\affiliation{APC, AstroParticule et Cosmologie,
Université de Paris, CNRS, Astroparticule et Cosmologie, F-75013 Paris, France}
\affiliation{Moscow Institute of Physics and Technology, Dolgoprudny, Moscow region, Russia}

\author{John Baker}
\affiliation{Gravitational Astrophysics Laboratory, NASA Goddard Space Flight Center, 8800 Greenbelt Rd., Greenbelt, MD 20771, USA}

\author{Tito Dal Canton}
\affiliation{Universit\'e Paris-Saclay, CNRS/IN2P3, IJCLab, 91405 Orsay, France}

 \begin{abstract}

Stellar-mass black hole binaries, like those currently being detected with the ground-based gravitational wave (GW) observatories LIGO and Virgo, are also an anticipated GW source in the Laser Interferometer Space Antenna (LISA) band.
LISA will observe them during the early inspiral stage of evolution; some of them will chirp through the LISA band and reappear some time later in the band of third generation ground-based GW detectors. Stellar-mass black hole binaries could serve as laboratories for testing the theory of general relativity and inferring the astrophysical properties of the underlying population. In this study, we assess LISA's ability to infer the parameters of those systems, a crucial first step in understanding and interpreting the observation of those binaries and their use in fundamental physics and astrophysics. We simulate LISA observations for several fiducial sources, setting the noise realization to zero, and perform a full Bayesian analysis. We demonstrate and explain degeneracies in the parameters of some systems. We show that the redshifted chirp mass and the sky location are always very well determined, with typical errors below $10^{-4}$ (fractional) and $0.4 \ {\rm deg^2}$. The luminosity distance to the source is typically measured within $40$-$60\%$, resulting in a measurement of the chirp mass in the source frame of $\mathcal{O}(1 \%)$. The error on the time to coalescence improves from $\mathcal{O}(1 \ {\rm day})$ to $\mathcal{O}(30 \ {\rm s})$ as we observe the systems closer to their merger.
We introduce an augmented Fisher-matrix analysis which gives reliable predictions for the intrinsic parameters compared to the full Bayesian analysis. Finally, we show that combining the use of the long-wavelength approximation for the LISA instrumental response together with the introduction of a degradation function at high frequencies yields reliable results for the posterior distribution when used self-consistently, but not in the analysis of real LISA data.

% For this purpose we present the first simulations of Bayesian inference on the parameters of stellar-mass black hole binaries.
%We scan the parameter space of SBHBs, simulating LISA observations and Bayesian inferences, in order to understand the features of the parameter estimation of SBHBs with LISA.
%We find large degeneracies between intrinsic parameters of the source, leading to large errors on individual masses and spins. Moreover, recovered distributions might peak sensibly away from the true value due to the effect of the prior. However, these degeneracies are partially broken for chirping systems allowing to measure individual masses and the specific combination of spins appearing at 1.5 PN order.
%The
%

 \end{abstract}

 \maketitle

 \section{Introduction}

Debuting with the first detection of a stellar-mass black hole binary (SBHB) in 2015 \cite{Abbott:2016blz}, the LIGO/Virgo collaboration has issued the first catalog of the gravitational wave (GW) sources identified during the first and second observing runs (O1 and O2) \cite{LIGOScientific:2018mvr,Abbott:2017oio,Abbott:2017gyy,TheLIGOScientific:2016pea,TheLIGOScientific:2017qsa,Abbott:2017vtc} and four noteworthy detections from the third observing run (O3) \cite{LIGOScientific:2020stg,Abbott:2020uma,GW190814,Abbott:2020tfl}. These observations of GWs in the $10$--$1000$ Hz band, which include 11 SBHBs mergers and two binary neutron star (BNS) mergers, have inaugurated the era of GW astronomy and opened a new window to the Universe, allowing to infer for the first time the properties of the population of compact binaries \cite{LIGOScientific:2018jsj,Abbott:2016ymx} and providing new tests of general relativity (GR) \cite{LIGOScientific:2019fpa,TheLIGOScientific:2016src,Abbott:2018lct}.

% Since the first gravitational waves (GWs) detection of a stellar mass  in 2015 \cite{Abbott:2016blz}, the LIGO/Virgo collaboration has announced ten additional observations during the first and second observing runs (O1 and O2) \cite{LIGOScientific:2018mvr,Abbott:2017oio,Abbott:2017gyy,TheLIGOScientific:2016pea,TheLIGOScientific:2017qsa,Abbott:2017vtc}, for a total of ten stellar-mass black hole binaries (SBHBs) and one binary neutron star (BNS), inaugurating the era of GW astronomy.
 %More observations from the third observing run (O3) are to be announced.

The Laser Interferometer Space Antenna (LISA) \cite{AHO17}, scheduled for launch in 2034, will observe GWs in a different frequency band (the $\rm mHz$ band) and, therefore, complement ground-based detectors. % , opening yet another window on the universe.
The strongest anticipated GW sources in the LISA data will be massive black hole binaries (MBHBs), with total mass in the range $10^4$--$10^7 M_{\odot}$ \cite{Klein:2015hvg}, and galactic white dwarf binaries (GBs). The latter are so numerous that they will
form a stochastic foreground signal dominating over instrumental noise in addition to a smaller number ($\sim 10^4$) of individually resolvable binaries \cite{Korol:2017qcx}. SBHBs with a total mass as large as those observed by LIGO and Virgo could also be detected by LISA during their early inspiral phase, long before entering the frequency band of ground-based detectors and merging \cite{Sesana:2016ljz}.
SBHBs in the ${\rm mHz}$ band can be at very different stages of their evolution, ranging from almost monochromatic sources to chirping sources which leave the LISA band during the mission \cite{Sesana:2016ljz}. We focus on resolvable SBHBs, one of the best candidates for multiband observations \cite{Sesana:2016ljz,AmaroSeoane:2009ui}.

Although SBHBs are not LISA's main target, the scientific potential of multiband observations with LISA and ground-based detectors is considerable. These could be used to probe low-frequency modifications due to deviations from GR \cite{Toubiana:2020vtf,Sesana:2016ljz,Gnocchi:2019jzp,Carson:2019rda,Vitale:2016rfr,Barausse:2016eii,Chamberlain:2017fjl} or to environmental effects \cite{Caputo:2020irr,Barausse:2014tra,Barausse:2014pra,Tamanini:2019usx,Cardoso:2019rou}, to facilitate electromagnetic follow up observations \cite{Caputo:2020irr}, or simply to improve parameter estimation (PE) over what is possible with ground-based interferometers alone \cite{Sesana:2016ljz,Vitale:2016rfr}. More precise measurements would improve the testing of competing astrophysical formation models.
Several scenarios have been suggested for the formation of SBHBs, such as stellar evolution of field binaries versus dynamical formation channels \cite{Postnov:2014tza,Benacquista:2011kv}. Moreover, the possibility that these black holes (BHs) are of primordial origin~\cite{Bird:2016dcv} cannot be completely discarded.
The various possible formation channels typically predict different distributions for the parameters of SBHBs, especially eccentricity and spin orientation/magnitude~\cite{Antonini:2012ad,Samsing:2017xmd,Gerosa:2018wbw} providing discriminating power in astrophysical model selection. A recent study has suggested that already a few ``special'' events, binaries with high primary mass and/or spin, would have a huge discriminating power \cite{Baibhav:2020xdf}. The specific impact of observations of SBHBs with LISA on astrophysical inference was considered in \cite{Gerosa:2019dbe,Samsing:2018isx,Nishizawa:2016jji,Nishizawa:2016eza,Breivik:2016ddj}.

LISA will observe MBHBs somewhere between a few days and few months before their merger, i.e.,~in their final, rapidly evolving inspirals \cite{Klein:2015hvg}. On the contrary, GBs are slowly evolving, almost monochromatic sources, and they will remain in the band during the whole LISA mission \cite{Timpano:2005gm}.
Resolvable SBHB signals will fall in between these two behaviors: they are long lived sources but are not monochromatic and
some SBHBs can chirp and leave the LISA band. In addition, all resolvable SBHBs are clustered at the high end of LISA's sensitive band. 
Thus, SBHBs will produce very peculiar signals of great diversity. In this work we do not address the question of
how to detect those sources, although it has been argued to be challenging~\cite{Moore:2019pke}. Instead, we assume that we have at our disposal an efficient detection method, and we focus on inferring the parameters of the detected signals. We also consider only one signal at a time, while in reality the SBHB signals will be superposed in the LISA data stream. The PE study presented here will also be valuable in building search tools
as we discuss at the end of the paper.

Most previous studies of PE for SBHBs with LISA relied on the Fisher approach, and used simple approximations to LISA's response to GW signals. While a quick and efficient method for forecasting studies, the Fisher approach might not be suited for the systems with low signal to noise ratio (SNR) and non-Gaussian parameter distributions \cite{Vallisneri:2007ev}.
In addition, SBHBs signals are long lived and emit at wavelengths comparable to LISA's size.
As a result, the commonly used long-wavelength approximation, also called low-frequency approximation \cite{Cutler:1997ta}, might not hold and could seriously bias the PE \cite{Vecchio:2004vt,Vecchio:2004ec}. In this work we thus consider the full LISA response as described in \cite{Marsat:2018oam, Marsat:2020rtl} and perform a full Bayesian analysis in zero noise of all the systems we consider. We also provide a comparison to Fisher-matrix-based PE and briefly comment on the impact of using the long-wavelength approximation.

Despite the on-going effort to infer the astrophysical formation channel of the SBHBs observed by LIGO/Virgo,
a huge uncertainty still remains. The situation will improve as we detect more signals. We expect that third generation detectors (Einstein Telescope \cite{Hild_2011,Punturo:2010zz,Ballmer:2015mvn}, Cosmic Explorer \cite{Abbott_2017_2}) will be operational in parallel with LISA, with SNR figures reaching hundreds or thousands, thus significantly improving the PE over current observations.
%  of the detected binaries should be greatly improved and so should be the inference on astrophysical formation channels.
Given the current uncertainty in the population of SBHBs, we cannot reliably specify
the properties of most detectable sources \cite{Sesana:2016ljz,Samsing:2018nxk,Kremer:2018cir,Wong:2018uwb,Gerosa:2019dbe,Kyutoku:2016ppx,Moore:2019pke,Gupta:2020lxa}.
In our study we focus on a fiducial system consistent with the population of currently observed SBHBs, instead of working with a randomized catalog of sources. We then perform a systematic scan of the parameter space by varying a few parameters at a time, investigating their qualitative impacts on the PE. We consider quasicircular binaries consisting of spinning BHs, with the spins aligned or antialigned with the orbital angular momentum, merging no later than twenty years from the beginning of observations. We start with a GW150914-like system \cite{Abbott:2016blz} and explore the parameter space by varying at most three parameters at a time.
For each of these systems we infer the posterior distribution, and we discuss the correlations between the parameters and the accuracy in measuring each parameter across the parameter space.

%Astrophysical formation models of SBHBs are very useful to explore different scenarios and provide signatures to distinguish between those. Nevertheless, even though recent models have been calibrated with GW observations, huge uncertainties remain due to little information on astrophysical mechanisms such as the common envelope phase \cite{1976IAUS...73...75P,Ivanova:2012vx} and the small number of GW detections.
% In particular, it should be noted that systems in the first catalogue released by the LIGO/Virgo collaboration \cite{LIGOScientific:2018mvr} are in average heavier than what was expected from early SBHBHs formation models. Since our aim is to characterise the PE acrros the parameter space of SBHBs we put aside the astrophysical motivation for the systems we study and the feasibilty of detection, of course still working under reasonable assumptions. To ensure this is the case
%Furthermore, it is not yet clear what SNR threshold will be required for detecting SBHBs with LISA, so the properties of the detectable population are very uncertain \cite{Sesana:2016ljz,Samsing:2018nxk,Kremer:2018cir,Wong:2018uwb,Gerosa:2019dbe,Kyutoku:2016ppx,Moore:2019pke}. For these reasons, instead of working on a catalogue of sources we perform a systematic scan of the parameter space.

The paper is organized as follows. In Sec.~\ref{analysis} we describe how we generate GW signals and our tools to perform PE. Then we give details on how we choose all the systems on which we perform PE in Sec.~\ref{setups}. Detailed description and analysis of PE results are given in Sec.~\ref{results}. There we also provide a comparison to a slightly modified version of Fisher matrix analysis and assess the validity of the long-wavelength approximation for the PE. Finally, in Sec.~\ref{ccl} we discuss the scientific opportunities offered by LISA observations of SBHBs in light of our results.

\section{Analysis method}\label{analysis}

\subsection{Bayesian framework}

Data measured by LISA ($d$) will consist in a superposition of GW signals ($s$) and a noise realization ($n$): $d=s+n$.
The instantaneous amplitude of a GW signal is much lower than noise, making its detection very challenging.
We use matched filtering as a main technique for detection, the main idea is to search for
a specific pattern (GW template) in the data \cite{Allen:2005fk}. It is done by correlating the data with a set of
GW templates in frequency domain ($\tilde{h}(f, \theta)$ which are functions of source parameters $\theta$). This correlation is given by the matched-filter overlap
\begin{equation}
	(d|h) = 4 {\mathcal Re} \left(\int_0^{+\infty} \frac{\tilde{d}(f) \ \tilde{h}^*(f)}{S_n(f)} {\rm d}f \right) \label{inner_product},
\end{equation}
where $S_n(f)$ is the power spectral density (PSD) of the detector noise, assumed to be stationary. In this work, we use the LISA ``proposal'' noise model given in \cite{AHO17}. We do not discuss the detectability of SBHBs in this paper; we assume that all sources discussed here can be detected, and we focus on the parameter extraction/estimation. We should note that the detection itself could be a challenge, at least for the traditional method of template banks~\cite{Moore:2019pke}, and that some mergers might only be detectable retroactively after being discovered by third generation ground based detectors \cite{Wong:2018uwb,Gupta:2020lxa,Ewing:2020brd}.

%  It indicates the level of noise at a given frequency and thus weights the integral giving the correlation. Indeed frequency bands where the sensitivity of the detector is higher (i.e. $S_n(f)$ is smaller) will give a higher contribution to $\rho$, reflecting the fact that in this frequency band we have a better ability to distinguish signal from noise.
%\

%The SNR of a GW signal is:
%
%\begin{equation}
% SNR(s)=\sqrt{(s|s)}.
%\end{equation}
%
%The higher the SNR, the higher the probability of the signal being detected. First studies on the possibility of observing SBHBs with LISA assumed a threshold SNR for detectability of $8$ \cite{Sesana:2016ljz,Wong:2018uwb,Gerosa:2019dbe,Kyutoku:2016ppx}. A recent work showed that, if using matched filtering only, the threshold would have to be increased up to $15$ for a direct detection with LISA, reducing drastically the number of events detectable by LISA \cite{Moore:2019pke}. But as pointed out in the same paper, new search methods might be developed which would allow to lower the threshold. We will then keep $8$ as a threshold. Moreover, in this work we assume search has been done and focus on PE.

We work in a Bayesian framework for the PE, treating the set of parameters of the source, $\theta$, as random variables. The Bayes theorem tells us that given the observed data $d$, the posterior distribution $p(\theta|d)$ is given by:

\begin{equation}
	p(\theta|d)=\frac{p(d|\theta)p(\theta)}{p(d)} \label{bayes} \,.
\end{equation}
 On the right hand side of this equation, $p(d|\theta)$ is the likelihood, $p(\theta)$ is the prior distribution and $p(d)$ is the evidence.
%The error on the parameters of the source and the correlations between them are linked to the shape of the posterior distribution.
As the noise and GW signal models will be fixed in this study, the evidence can be seen as a normalization constant that does not need an explicit calculation. Assuming noise to be stationary and Gaussian, the likelihood is given by:
\begin{equation}
	\mathcal{L} = p(d|\theta) = \exp \left[ -\frac{1}{2} (d-h(\theta)|d-h(\theta)) \right] \,.
\end{equation}

In order to speed up the computation, we set the noise realization to zero as $n=0$, so that $d=s$. The addition of noise to the GW signal is not expected to drastically affect the PE, leading at most to a displacement of the centroid of the posterior distribution within the confidence intervals (CI) (with the probability defined by CI). Thus, the analysis of the posterior distribution itself should remain representative in the presence of noise (still assuming Gaussianity). We consider only one source at a time and we neglect all possible systematic errors due to signal mismodeling: $s=h(\theta_0)$ with $\theta_{0}$ the parameters of the GW source.
% We consider only GW signal at once and neglect the inaccuracies of GW templates in reproducing the true GW signal emitted by the sources so $s=h(\theta_0)$.
% Our analysis consists in generating an injection signal for a given source $\theta_{0}$ and computing the posterior distribution of the source parameters $\theta$ using \ref{bayes}.
 Under these simplifications, the log-likelihood is given by (up to a normalization constant in $\mathcal{L}$):
 \begin{equation}
 	\log \mathcal{L}=-\frac{1}{2}(h(\theta_{0})-h(\theta)|h(\theta_0)-h(\theta)) \,. \label{loglike}
 \end{equation}
 
 \begin{table}
  \begin{center}
   \begin{tabular}{|c|c|c|}
   
   \hline
   
  Symbol & Meaning  &  Expression \\
   \hline

  $m_1$ & Mass of the primary BH & / \\

  $m_2$ & Mass of the secondary BH & / \\

   $\chirpm$ & Chirp mass & $\chirpm=\left ( \frac{m_1^{3}m_2^{3}}{m_1+m_2} \right )^{1/5}$ \\

  $\eta$ & Symmetric mass ratio & $\eta=\frac{m_1m_2}{(m_1+m2)^2}$   \\

  $q$ & Mass ratio & $q=\frac{m_1}{m_2}$   \\
 
  $M$ & Total mass & $M=m_1+m_2=\chirpm \eta^{-3/5}$   \\

\multirow{2}{*}{$\chi_1$} & Spin of the primary BH along  & \multirow{2}{*}{/} \\
 & the orbital angular momentum & \\

 \multirow{2}{*}{$\chi_2$} & Spin of the secondary BH along & \multirow{2}{*}{/}  \\
 & the orbital angular momentum & \\
  
   $\chi_+$ & Effective spin & $\chi_+=\frac{m_1\chi_1+m_2\chi_2}{m_1+m_2}$  \\

  $\chi_-$ & Antisymmetric spin combination & $\chi_+=\frac{m_1\chi_1-m_2\chi_2}{m_1+m_2}$  \\

  \multirow{2}{*}{$\chi_{\rm PN}$} &  \multirow{2}{*}{1.5 PN spin combination} &  $\chi_{\rm PN}=\frac{\eta}{113} [ (113q+75)\chi_1 $  \\

 & & $+(\frac{113}{q}+75)\chi_2 ]$ \\
 
 $f_0$ & Initial frequency  & / \\
 
 $t_c$ & Time to coalescence  & / \\ % Eq.~\eqref{eq:tf} \\

  $\lambda$ & Longitude in the SSB frame & /  \\

   $\beta$ & Latitude in the SSB frame & /   \\

   $\psi$ & Polarization angle  & /  \\

   $\varphi$ & Initial phase & /  \\

   $\iota$ &  Inclination & / \\

  $D_L$ & Luminosity distance in Mpc & / \\

$z$  & Redshift & $z(D_L)$ \cite{Aghanim:2018eyx}  \\
  \hline

   \end{tabular}
   \end{center}
    \caption{Parameters used throughout the paper and their explicit expressions when necessary.}\label{acronyms}
  \end{table}

Since LISA will only observe the inspiral phase of these binaries, we expect that the dominant 22 mode is sufficient and we neglect the contribution of all other subdominant harmonics. We use the model called PhenomD \cite{Husa:2015iqa,Khan:2015jqa} to generate $\tilde{h}_{2,\pm2}$ and compute the LISA response to generate the time delay interferometry (TDI) observables $A$, $E$, and $T$ (see, e.g., \cite{Tinto:2004wu}) as described in~\cite{Marsat:2018oam, Marsat:2020rtl}. The three TDI observables constitute independent datasets, therefore the log-likelihood is actually a sum of three terms like Eq.~\eqref{loglike}, one per TDI observable.

Similarly to the treatment of galactic binaries, we parametrize the sources by their initial frequency and phase at the start of the observation, instead of the time to coalescence and phase at coalescence (more suitable in LIGO/Virgo data analysis or for MBHBs with LISA). We define the initial time as the moment LISA starts observing the system.
A system is characterized by (i) five intrinsic parameters: the masses ($m_1$ and $m_2$), the GW frequency at which LISA starts observing the system ($f_0$) and spins ($\chi_1$ and $\chi_2$); and (ii) six extrinsic parameters: the position in the sky defined in the solar system barycenter frame (SSB) ($\lambda$ and $\beta$), the polarization angle ($\psi$), the azimuthal angle
of the observer in the source frame ($\varphi$), the inclination of the orbital angular momentum with respect to the line of sight ($\iota$) and the luminosity distance to the source ($D_L$). We need only two out of the six general spin parameters to describe the system because we assume each spin to be aligned (or antialigned) with the orbital angular momentum.
%$\varphi$ is defined as the angle of the rotation performed on the observer such that it is in the direction of the GW. With this definition $\varphi$ picks up a minus sign relative to the usual convention.

We introduce a set of sampling parameters for which we expect the posterior distribution to be a simple function, i.e.~close to either a uniform or Gaussian distribution, based on the properties of post-Newtonian (PN) inspiral waveforms \cite{Blanchet:2002av,Buonanno:2006ui,Buonanno:2009zt}.
These parameters are $\theta=(\chirpm, \eta, f_0, \chi_+, \chi_-, \lambda, \sin(\beta), \psi, \varphi, \cos (\iota), \log_{10}(D_L))$, where $\chirpm=\frac{m_1^{3/5}m_2^{3/5}}{(m_1+m_2)^{1/5}}$ is the chirp mass, $\eta=\frac{q}{(1+q)^2}$ is the symmetric mass ratio with $q=\frac{m_1}{m_2} > 1$ being the mass ratio, $\chi_+=\frac{m_1 \chi_1+ m_2 \chi_2}{m_1+m_2}$ is the effective spin (often denoted $\chi_{\rm eff}$ in the literature) and $\chi_-=\frac{m_1 \chi_1-m_2 \chi_2}{m_1+m_2}$ is an antisymmetric spin combination. For easier reference, in Table \ref{acronyms} we list the parameters used throughout this paper, together with their explicit expressions. Some of these combinations will be introduced later in the paper.

For the simulated data we assume a sampling rate of $1 \ {\rm Hz}$ so the Nyquist frequency is $f_{\rm Ny}=0.5 \ {\rm Hz}$. When computing inner products given by Eq.~\eqref{inner_product}, we generate templates from $f_0$ up to $f_{\rm max} = {\rm min}(f_{\rm Ny},f_{T_{\rm obs}}$) where $f_{T_{\rm obs}}$ is the frequency reached by the system after the observation time $T_{\rm obs}$. We consider two mission durations: $T_{\rm obs}=4$ yr and $T_{\rm obs}=10$ yr.
Details on fast LISA response generation and likelihood computation are given in \cite{Marsat:2020rtl}.

Due to the high dimensionality of the problem, we need an efficient way to explore the parameter space. We do this by means of a Markov chain Monte Carlo (MCMC) algorithm \cite{Karandikar2006}. More specifically we designed a Metropolis-Hastings MCMC (MHMCMC) \cite{10.2307/2684568} for this purpose that we present next. A less costly
 alternative would be to use the PE based on the Fisher information matrix. We will show how one can modify the Fisher matrix to make it a robust PE tool in the following subsections. We exploit the metric interpretation of Fisher matrix in the MHMCMC, thus, we start by reviewing some basics on the Fisher matrix approach and delegate comparison with Bayesian PE to Sec.~\ref{results_fm}.

\subsection{Fisher matrix}\label{fisher_mat}

In the Fisher matrix approach, the likelihood is approximated by a multivariate Gaussian distribution \cite{Vallisneri:2007ev}:
\begin{equation}
	p(d|\theta) \propto e^{-\frac{1}{2}F_{ij}(\theta_{0})\Delta \theta_i \Delta \theta_j} \,,
\end{equation}
where $\Delta \theta=\theta-\theta_{0}$ and $F$ is the Fisher matrix given by:
\begin{equation}
	F_{ij}(\theta)= \left . (\partial_i h | \partial_j h) \right |_{\theta} \label{def_fisher} \,,
\end{equation}
where the partial derivative $\partial_i$ denotes the derivative with respect to $\theta_i$. Similarly to the likelihood, we actually have a sum of three terms, one for each TDI observable. We assume this to be implicit in the following.
The inverse of $F$ is the Gaussian covariance matrix of the parameters, which gives an estimate of the error on each parameter. The Fisher approach has been extensively used in the studies of LISA's scientific capability, thanks to its simplicity. However, for systems with low SNR as the ones we consider, the Fisher approximation might not be valid \cite{Vallisneri:2007ev} and we need to perform a full Bayesian analysis.
%In order to increase numerical stability we split the phase and amplitude contribution to the waveform derivative, noting $\tilde{h}(\theta)=Ae^{i\psi}$ we get:

%\begin{equation}
% F_{ij}(\theta)= 4 {\mathcal Re} \left(\int_0^{+\infty} \frac{(\partial_i A+iA\partial_i\Psi )(\partial_j A-iA\partial_j\Psi )}{S_n(f)} {\rm d}f \right)
%\end{equation}

The Fisher matrix has an alternative interpretation: it can be seen as a metric on the parameter space associated with the distance defined by the inner product \eqref{inner_product}:
\begin{align}
	||h(\theta+\delta \theta)-h(\theta)||^2 = &(h(\theta+\delta \theta)-h(\theta)|h(\theta+\delta \theta) -h(\theta)) \nonumber\\
 	\simeq& \left . (\partial_i h \  \delta \theta^i | \partial_j h \ \delta \theta^j) \right |_{\theta}  \nonumber\\
 	=& F_{ij}(\theta) \delta \theta^i \delta \theta^j.
\end{align}
We exploit this property in our MHMCMC sampler.

%We use a multivariate Gaussian distribution centred around $0$ with standard deviation $\sigma$ for $\chi_+$ and $\chi_-$ to approximate the prior. Under this approximation

\subsection{Metropolis-Hastings MCMC}\label{mhmcmc}

We sample the \emph{target} distribution $p(\theta|d)$ by means of a Markov chain, generated from a transition function $P(\theta,\theta')$ satisfying the detailed balance condition:
\begin{equation}
 p(\theta|d)P(\theta,\theta')=p(\theta'|d)P(\theta',\theta).
\end{equation}
We build the transition function from a proposal function $\pi$ such that $P(\theta,\theta')=\pi(\theta,\theta')a(\theta,\theta')$ where $a(\theta,\theta')$ is the acceptance ratio defined as:
\begin{equation}
 a(\theta,\theta') =  \left \{
\begin{array}{ll}
 \mathrm{min} \left( 1,\frac{p(\theta'|d)\pi(\theta',\theta)}{p(\theta|d)\pi(\theta,\theta')} \right) \ {\rm if}  \ \pi(\theta,\theta') \neq 0  \\
    0 \ {\rm otherwise.}
  \end{array}
  \right. \label{acc_ratio}
\end{equation}
It is easy to verify that $P$ satisfies the detailed balance condition for any choice of $\pi$.
In practice, a jump from a point $\theta$ to $\theta'$ is proposed using the function $\pi(\theta,\theta')$ and the new point is accepted with probability $a(\theta,\theta')$. If the point is not accepted, the chain remains at $\theta$. In both cases, we update the current state of the chain. By repeating this procedure, we obtain a sequence of samples of the target distribution. We see from the expression of the acceptance ratio that points with higher posterior density are more likely to be accepted, thus the chain will tend to move towards regions of higher posterior density, exploring all regions of the parameter space compatible with the observed data.
In theory, the chain should converge regardless of the proposal function and starting point of the chain, but in practice it may take an inconveniently large time to do so unless the proposal and starting points are chosen wisely. Since we are interested in high posterior regions, we start the chain from the true signal parameters, i.e.~the maximum-likelihood point. The maximum-likelihood point coincides with the maximum posterior point if all priors are flat, but this is not true in general and the maximum posterior point can depend on the adopted prior distribution.
% Although generically the maximum likelihood point does not coincide with the maximum posterior point, unless the prior has no support in that region it will still be in a region of high posterior.
 Even though the posterior does not depend on the proposal, the convergence, efficiency and resolution of tails of the distribution do very strongly depend on the particular choice; ideally, an efficient proposal should closely resemble the target posterior distribution.
Thus, most of the work goes into building an efficient proposal function. %We describe our choice of proposals  shortly.
Note that for a symmetric proposal ($\pi(\theta,\theta')=\pi(\theta',\theta)$), the acceptance ratio is simply given by the ratio of the posterior distributions. This specific case is called Metropolis MCMC \cite{Metropolis:1953am} and is the one we consider.
%We build the transition function we
%By taking a proposal function $\pi(\theta,\theta')$ and defining an acceptance ratio $a(\theta,\theta')$ as:

%The transition function is is obtained from a proposal function such that
%proposal distribution ($\pi$) satisfying the balance equation:  \st{This is not balance equation. It is just a symmetric proposal which by definition satisfies the balance equation. Not that MCMC with the symmetric proposal is actually Metropolis MCMC. Hastings introduced extra factors for non-symmetric proposals}
%By doing so we accumulate samples forming the shape of the posterior distribution.

Our runs are done in two steps: we first run a short MCMC chain ($\simeq 10^5$ points) to explore the parameter space and then use the covariance matrix of the points obtained from this chain to build a multivariate Gaussian proposal that we use in a longer chain.
During the first stage, called burn-in, we use a block diagonal covariance matrix.
We split the set of parameters in three groups: intrinsic parameters ($\chirpm$, $\eta$, $f_0$, $\chi_+$, $\chi_-$), angles except the inclination ($\lambda$, $\sin(\beta)$, $\psi$, $\varphi$) and inclination distance ($\cos (\iota)$, $\log_{10}(D_L)$). Each block is computed inverting the Fisher matrix of that group of parameters.
This separation was based on the intuition, well verified in practice, that the stronger correlations are within these groups of parameters and is intended to avoid numerical instabilities that may arise when dealing with full Fisher matrices. Note that by making this choice we do not discard possible correlations between parameters of different groups; we are simply not taking them into account when proposing points based on the Fisher matrix. If those correlations exist, they should appear in the resulting covariance matrix that we use to build a proposal for the main chain. Failing to include existing correlations could reduce the efficiency of our sampler in its exploratory, or burn-in, phase; however the splitting can easily be adapted if needed.

We rotate the current state vector $\theta$ to the basis of the covariance matrix's eigenvectors. In this basis the covariance matrix is diagonal, formed by the eigenvalues of the covariance matrix in the original basis. Because for some parameters the distribution is very flat, the eigenvalues of the covariance matrix predicted by Fisher can be very large, reducing the efficiency of the sampler. This is usually the case for poorly constrained but bounded parameters like $\cos (\iota)$ and spins. To avoid this issue, we truncate the eigenvalues of the ($\cos (\iota)$,$\log_{10}(D_L)$) matrices and define an effective Fisher matrix accounting for the finite extent of the prior on spins: $F_{\rm eff} = F + F^{\rm p}$. We take $F^{\rm p}_{\chi_+,\chi_+} = F^{\rm p}_{\chi_-,\chi_-} = \frac{1}{\sigma^2}$ with $\sigma=0.5$. We motivate this choice in Sec.~\ref{results_fm}.

In order to improve the sampling efficiency in the event of complicated correlations between intrinsic parameters, we exploit the metric interpretation of Fisher matrix and occasionally recompute the covariance matrix for the first group of parameters with a given probability. By doing so we might violate the balance equation, but this is only done during the burn-in stage (exploration of the parameter space); the resulting points are then discarded from the analysis.

We test the convergence of the chains by running multiple chains with different random number generator seeds, checking that they all give similar distributions and computing the Gelman-Rubin diagnosis \cite{Gelman:1992zz} for all the parameters. Potential scale reduction factors below 1.2, as the ones we get, indicate that the chains converged \cite{Gelman:1992zz}. For each chain we accumulate $10^3$--$10^4 $ independent samples (by thinning the full chain by the autocorrelation length) which takes $4$--$7$ hours on a single CPU thanks to the fast likelihood computation and LISA response generation presented in \cite{Marsat:2020rtl}.

\section{Setups}\label{setups}

\subsection{Systems}

We start by considering a system with masses and spins compatible with the first detected GW signal (GW150914) and label it \emph{Fiducial} system. Its parameters are given in Table \ref{params_fiducial} along with its SNR assuming LISA mission lifetimes $T_{\rm obs}=4\mathrm{yr}$ and $T_{\rm obs}=10\mathrm{yr}$. We give both the detector and source frame masses, related by $m=(1+z)m_{s}$ where $z$ is the cosmological redshift and subscripts $s$ denote parameters in the source frame. We adopt the cosmology reported by the Planck mission (2018) \cite{Aghanim:2018eyx}. Note that $T_{\rm obs}$ is the mission duration, not the time spent by the system in the LISA band and we assume an ideal 100\% duty cycle. The initial frequency is derived from the time to coalescence from the beginning of LISA observation ($t_c$) that we fix to eight years for the \emph{Fiducial} system. Thus, with $T_{\rm obs} = 4$ yr the \emph{Fiducial} system is observed for a fraction of its inspiral, while with $T_{\rm obs} = 10$ yr the same system is observed for eight years before exiting the LISA band and coalescing. The sky location is given in the SSB frame.
In the following, subscripts $f$ refers to the \emph{Fiducial} system.

We explore the parameter space of SBHBs by changing a few parameters of the \emph{Fiducial} system at a time.
We list all the systems we consider in the following subsections, specifying what are the changes with respect to the \emph{Fiducial} system and the corresponding labels.
For all systems we consider the two possible mission durations quoted above, unless another choice is specified.

In Table \ref{systems} we show the considered systems and their respective SNR. Note that we chose to use the LISA proposal noise level \cite{AHO17}, which does not include a 50\% margin introduced to form the ``science requirements'' SciRDv1~\cite{scirdv1}. The SNRs would thus be significantly lower with SciRDv1. From the point of view of the PE, using one or the other noise model amounts to a constant rescaling of the noise PSD $S_{n}$, with the same effect as rescaling the distance to the source.

 \begin{table}
  \begin{center}
   \begin{tabular}{|c|c|c|}

   \hline

  $m_1$ (M\textsubscript{\(\odot\)}) & \multicolumn{2}{|c|}{$40$} \\

  \hline

  $m_2$ (M\textsubscript{\(\odot\)}) & \multicolumn{2}{|c|}{$30$}  \\

  \hline

   $m_{1,s}$ (M\textsubscript{\(\odot\)}) & \multicolumn{2}{|c|}{$36.2$} \\

  \hline

  $m_{2,s}$ (M\textsubscript{\(\odot\)}) & \multicolumn{2}{|c|}{$27.2$}  \\

  \hline

  $t_c$ (yrs) & \multicolumn{2}{|c|}{$8$}   \\

\hline

$f_0$ (mHz) & \multicolumn{2}{|c|}{$12.7215835397$}  \\

\hline

  $\chi_1$ & \multicolumn{2}{|c|}{$0.6$} \\

  \hline

  $\chi_2$ & \multicolumn{2}{|c|}{$0.4$}  \\

  \hline

  $\lambda$ (rad) & \multicolumn{2}{|c|}{$1.9$}  \\

  \hline

   $\beta$ (rad) & \multicolumn{2}{|c|}{$\pi/3$}  \\

  \hline

   $\psi$ (rad) & \multicolumn{2}{|c|}{$1.2$}   \\

  \hline

   $\varphi$ (rad) & \multicolumn{2}{|c|}{$0.7$}\\

  \hline

   $\iota$ (rad) &  \multicolumn{2}{|c|}{$\pi/6$} \\

  \hline

  %$f_0$ in Hz & $0.0125 $& \\
  $D_L$ (Mpc) & \multicolumn{2}{|c|}{$250$} \\

  \hline

$z$  & \multicolumn{2}{|c|}{$0.054$} \\

  \hline

  $T_{\rm obs}$ (yrs) &  $4$ & $10$  \\

  \hline

  SNR &  $13.5$ & $21.5$ \\

  \hline

   \end{tabular}
   \end{center}
    \caption{Parameters of a representative SBHB system labeled \emph{Fiducial}. The masses and spins of this system are compatible with GW150914 \cite{Abbott:2016izl}. The initial frequency is computed such that the system is merging in eight years from the start of LISA observations. We consider two possible durations of the LISA mission: four and ten years (in the latter case, the signal stops after eight years at coalescence). Subscripts $s$ denote quantities in the source frame, bare quantities are in the detector frame. The sky location is given in the SSB frame.}\label{params_fiducial}
  \end{table}

\subsubsection{Intrinsic parameters}

Unless specified we take $t_c=8 \ {\rm years}$ and we compute the initial frequency corresponding to the chosen $t_c$.
Changing $t_c$ (or equivalently $f_0$) amounts in shifting the GW signal in frequency and also defines its frequency bandwidth (within the chosen observation time). We consider the following variations in the intrinsic parameters:
%stage of the binary evolution at which LISA starts observing it. Of course, during the actual LISA mission we will not be free to chose the initial frequencies of the observed systems.
    \begin{itemize}
     \item Time left to coalescence at the beginning of LISA observations:

     \emph{Earlier}: $t_c=20$ yr,

     \emph{Later}: $t_c=2$ yr

     \item Chirp mass keeping the mass ratio unchanged:

     \emph{Heavy}: $\chirpm =1.5 \mathcal{M}_{c,f}$, $q=q_f$, $D_L=445 \ {\rm Mpc}$

     \emph{Light} $\chirpm =\frac{\mathcal{M}_{c,f}}{1.5} $, $q=q_f$, $D_L=150 \ {\rm Mpc}$

     \item Mass ratio, keeping the chirp mass unchanged:

     \emph{q3}: $q=\frac{m_1}{m_2}=3$, $\chirpm =\mathcal{M}_{c,f}$

     \emph{q8}: $q=\frac{m_1}{m_2}=8$, $\chirpm =\mathcal{M}_{c,f}$

     \item Spins:

     \emph{SpinUp}: $\chi_1=0.95$, $\chi_2=0.95$

     \emph{SpinDown}:  $\chi_1=-0.95$, $\chi_2=-0.95$

     \emph{SpinOp12}: $\chi_1=0.95$, $\chi_2=-0.95$

     \emph{SpinOp21}:  $\chi_1=-0.95$, $\chi_2=0.95$.

     \end {itemize}

     For the \emph{Heavy} and \emph{Light} systems we scaled the distance so that the SNR remains the same as for the \emph{Fiducial} system in the case $T_{\rm obs}=10\mathrm{yr}$. Changing spins or mass ratio barely affects the SNR, so we do not change the distance for those systems. Since the \emph{Earlier} system merges in two years, increasing the observation time from four to ten years has no impact.

    \subsubsection{Extrinsic parameters}

Changes in extrinsic parameters do not affect the time to coalescence, so all systems below have the same initial frequency as the \emph{Fiducial} system. We consider the following variations in the extrinsic parameters (those depending on the relative orientation of the source to the observer):

     \begin{itemize}

     \item Sky location in the SSB frame:

     \emph{Polar}: $\beta=\frac{\pi}{2}-\frac{\pi}{36}$, $\lambda=\lambda_f$

     \emph{Equatorial}: $\beta=\frac{\pi}{36}$, $\lambda=\lambda_f$

     \item Inclination:

     \emph{Edgeon}: $\iota=\frac{\pi}{2}-\frac{\pi}{36}$, $D_L=150 \ {\rm Mpc}$, $T_{\rm obs}=10\mathrm{yr}$

     \item Distance:

     \emph{Close}: $D_L=190 \ {\rm Mpc}$, $T_{\rm obs}=4\mathrm{yr}$

     \emph{Far}: $D_L=350 \ {\rm Mpc}$, $T_{\rm obs}=10\mathrm{yr}$

     \emph{Very Far}: $D_L=500 \ {\rm Mpc}$, $T_{\rm obs}=10\mathrm{yr}$

    \end{itemize}

The drop in SNR being very large for an almost edge-on system, we decrease the distance of the \emph{Edgeon} system to maintain a reasonably high SNR. For the same reason, we use only $T_{\rm obs}=10\mathrm{yr}$ in this case.
The goal of the variation in distance is to assess the impact of the SNR on the PE, all other things being equal. This also mimics the effect of varying the noise level and the duty cycle. For the \emph{Close} system we only consider the $T_{\rm obs}=4\mathrm{yr}$ case and for the \emph{Far} and \emph{Very Far} systems we only consider the $T_{\rm obs}=10\mathrm{yr}$ case.

%The remaining changes concern the detector response and a speculative change in $f_{max,LISA}$:

%\begin{itemize}

 %\item Detector response:

% \emph{Low-f}: low frequency approximation

 %\item Maximum frequency:

 %\emph{fmax0.1}: $f_{max,LISA}=0.1 \ {\rm Hz}$

 %\emph{fmax0.05}: $f_{max,LISA}=0.05 \ {\rm Hz}$.

%\end{itemize}

%In the \emph{Low-f} case we use a simplified version of the LISA response: the low frquency (or long wavelength) approximation \cite{Cutler:1997ta}. In this limit finite arm size effects vanish and LISA response is similar to two LIGO/Virgo type detectors rotated from each other by $\pi/4$. As suggested in its name, this approximation is not expected to hold at the high frequencies at which we observe SBHBs. We compare how does the PE change for the \emph{Fiducial} system when using this approximation.
%As we will discuss in section \ref{results}, high frequencies play a very important role in the estimation of source parameters. We want to assess how much worst would the PE of the \emph{Fiducial} system get if $f_{max,LISA}$ is decreased. Since this only affects systems leaving the band during the mission we consider this modification only for the  $T_{\rm obs}=10 \ years$ case.

\begin{table}
  \begin{center}
   \begin{tabular}{c|c|c|}

    \cline{2-3}

    & $T_{\rm obs}=4\mathrm{yr}$ & $T_{\rm obs}=10\mathrm{yr}$ \\

    \hline

    \multicolumn{1}{|c|}{\emph{Fiducial}} & $13.5$ & $21.1$ \\

    \hline

    \multicolumn{1}{|c|}{\emph{Earlier}} & $10.3$ & $17.2$ \\

   \hline

   \multicolumn{1}{|c|}{\emph{Later}} & $11.8$ & / \\

   \hline

  \multicolumn{1}{|c|}{\emph{Heavy}} & $12.8$ & $20.9$   \\

  \hline

  \multicolumn{1}{|c|}{\emph{Light}} & $14.1$ & $21.1$  \\

  \hline

  \multicolumn{1}{|c|}{\emph{q3}} & $13.5$ & $21.1$  \\

\hline

 \multicolumn{1}{|c|}{\emph{q8}} & $13.5$ & $21.1$   \\

  \hline

  \multicolumn{1}{|c|}{\emph{SpinUp}} & $13.5$ & $21.1$  \\

\hline

 \multicolumn{1}{|c|}{\emph{SpinDown}} & $13.5$ &  $21.1$  \\

\hline

 \multicolumn{1}{|c|}{\emph{SpinOp12}} & $13.5$ & $21.1$  \\

\hline

 \multicolumn{1}{|c|}{\emph{SpinOp21}} & $13.5$ &  $21.1$  \\

\hline

  \multicolumn{1}{|c|}{\emph{Polar}} & $12.8$ & $20.1$  \\

  \hline

  \multicolumn{1}{|c|}{\emph{Equatorial}} & $14.9$ & $23.1$  \\

  \hline

  \multicolumn{1}{|c|}{\emph{Edgeon}} & / & $14.7$  \\

  \hline

  \multicolumn{1}{|c|}{\emph{Close}} & $17.8$ & /  \\

  \hline

  \multicolumn{1}{|c|}{\emph{Far}} & / & $15.1$    \\

  \hline

  \multicolumn{1}{|c|}{\emph{Very Far}} & / &$10.6$  \\

   \hline

   %5\multicolumn{1}{|c|}{\emph{Low-f}} & $15.0$ &$34.2$  \\

  %\hline

   %\multicolumn{1}{|c|}{\emph{fmax0.05}} & / &$20.7$  \\

  %\hline

  %\multicolumn{1}{|c|}{\emph{fmax0.1}} & / & $21.0$  \\

  %\hline

   \end{tabular}
   \end{center}
    \caption{SNR of all systems considered, computed with the LISA proposal noise level given in \cite{AHO17}. Different systems are derived from the \emph{Fiducial} system, varying a few parameters at once. We use the \emph{Full} response.}\label{systems}
  \end{table}

  \subsection{Prior}\label{priors}

Regarding the Bayesian analysis, we take our fiducial prior to be flat in $m_1$ and $m_2$ with $m_1 \geq m_2$, flat in spin magnitude between $-1$ and $1$, flat in initial frequency, volume uniform for the source location and flat in the source orientation, its polarization and its initial phase. For phase and polarization, since only $2\varphi$ (for a 22-mode waveform) and $2\psi$ intervene, we restrict to an interval of $\pi$. We obtain the prior probability density function (PDF) in terms of the sampling parameters by computing the Jacobian of the transformation from ($m_1,m_2,\chi_1,\chi_2,D_L$) to ($\chirpm,\eta,\chi_+,\chi_-,\log_{10}(D_L)$) which gives:
\begin{equation}
 p_f(\theta) =
	\begin{cases}
		& N \frac{\chirpm \eta^{-11/5}D_L^3}{\sqrt{1-4\eta}} \; {\rm if} \; 0.05 \leq \eta \leq 0.25 \,, \\
		& 0 \; {\rm otherwise.}
	\end{cases} \label{fid_prior}
\end{equation}
%\begin{equation}
% p_f(\theta) = \left \{
%\begin{array}{ll}
%N \frac{\chirpm \eta^{-11/5}D_L^3}{\sqrt{1-4\eta}} \ {\rm if}   \  -1\leq \chi_1, \chi_2, \cos \iota, \sin(\beta) \leq 1, \ \\
% \ \ \ \ \ \ \ \ \ \ \ \ \ \ \ \ \ \ \ \  0.05 \leq \eta \leq 0.25, \ |\lambda-\lambda_0| \leq \pi \ \\
% \ \ \ \ \ \ \ \ \ \ \ \ \ \ \ \ \ \ \ \ {\rm and} \ |\psi, \phi -\psi_0, \phi_0| \leq \pi/2  \\
%    0 \ {\rm otherwise.}
%  \end{array}
%  \right. \label{fid_prior}
%\end{equation}
Just like the evidence in Eq.~\eqref{bayes}, $N$ acts only as a normalization constant and thus it is of no importance for us. The lower limit for $\eta$ was set according to the maximum mass ratio up to which PhenomD is calibrated ($q=16$)~\cite{Husa:2015iqa,Khan:2015jqa}. The range of chirp mass, initial frequency and distance are orders of magnitude larger than the posterior support so they do not affect the posterior.
We label this prior as \emph{Flatphys} and use it by default unless we specify some other choice, for example, we will consider two additional priors:
\begin{itemize}

  \item \emph{Flatmag}: uniform prior for the spins orientation and magnitude

  \item \emph{Flatsampl}: flat prior in $\chirpm$, $\eta$ and $\log_{10}(D_L)$.

\end{itemize}
 In the \emph{Flatmag} case we start from a full 3D spin prior, uniform in [0,1] for the spins amplitude and uniform on the sphere for the spins orientation. We then consider only the spin projections on the orbital momentum, thus ignoring the in-plane spin components. The resulting prior is $p(\chi_i)=-\frac{1}{2}\ln(|\chi_i|)$. The \emph{Flatmag} PDF is: $p(\theta)=p_f(\theta)p(\chi_1)p(\chi_2)$ where $p_f(\theta)$ is given in Eq.~\eqref{fid_prior}. This is the prior generally used by the LIGO/Virgo collaboration \cite{LIGOScientific:2018mvr}.

The \emph{Flatsampl} PDF is given by:
\begin{equation}
 p(\theta) =
	\begin{cases}
		& N\frac{1}{\eta} \; {\rm if} \; 0.05 \leq \eta \leq 0.25 \,, \\
		& 0 \; {\rm otherwise.}
	\end{cases} \label{flatsampl_prior}
\end{equation}
%\begin{equation}
% p(\theta) = \left \{
%\begin{array}{ll}
%N\frac{1}{\eta} \ {\rm if }   \  -1\leq \chi_1, \chi_2, \cos \iota, \sin(\beta) \leq 1, \ \\
% \ \ \ \ \ \ \ \ \ \ \ \ \ \   0.05 \leq \eta \leq 0.25, \ |\lambda-\lambda_0| \leq \pi \ \\
% \ \ \ \ \ \ \ \ \ \ \ \ \ \ {\rm and} \ |\psi, \phi -\psi_0, \phi_0| \leq \pi/2  \\
%    0 \ {\rm otherwise}.
%  \end{array}
%  \right. \label{flatsampl_prior}
%\end{equation}
 This prior has no astrophysical motivation; we will use it to compare the Fisher-based PE to our full Bayesian inference in Sec.~\ref{results_fm}.

\begin{figure}
\centering
 \includegraphics[width=0.5\textwidth]{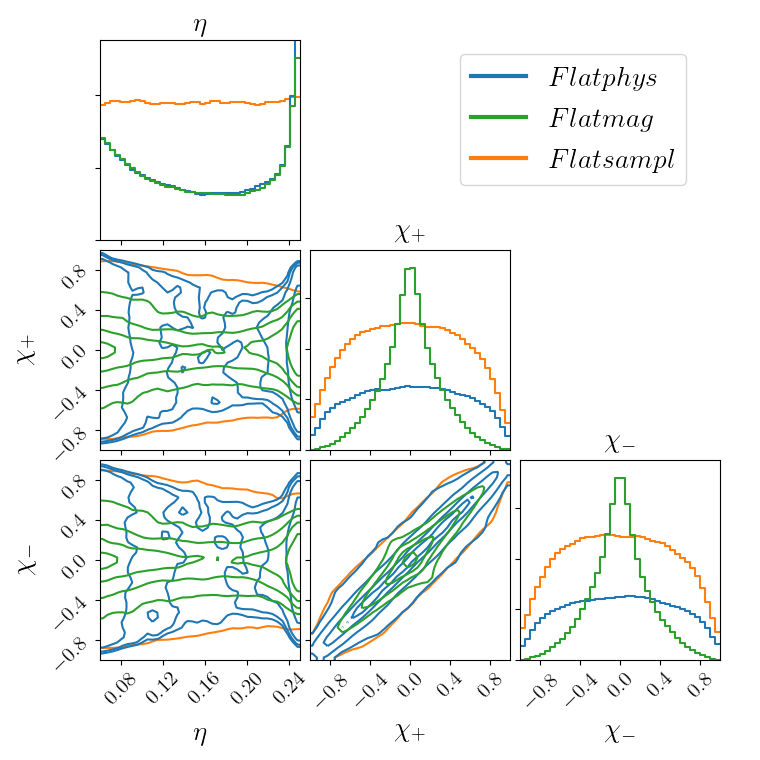}\\
 \centering
 \caption{Comparison between the \emph{Flatphys} (blue), \emph{Flatmag} (green) and \emph{Flatsampl} (orange) priors for $\eta$, $\chi_+$, and $\chi_-$.}\label{comp_priors}
\end{figure}

We find it instructive to illustrate how the nontrivial priors look like. As we will show later in Sec.~\ref{results}, the chirp mass can be constrained by Bayesian analysis to a fractional error of $10^{-4}$, so we can impose a narrow constraint on the prior. The chirp mass is nontrivially coupled to other parameters (as we will show in great details in the following sections), and constraining it to the narrow interval introduces nonlinear slicing in other parameters. Note that the imposed interval ($10^{-3}$ in relative terms) is still much broader than the typical measurement error. In Fig.~\ref{comp_priors} we display the \emph{Flatphys}, \emph{Flatmag} and \emph{Flatsampl} prior distributions for $\eta$, $\chi_+$, and $\chi_-$ obtained by restraining the chirp mass to the specified interval. The remarkable features of our fiducial prior, the \emph{Flatphys} prior, are the double peak at $\eta=0.25$ and $\eta=0$ and the bell-like shape for the $\chi_+$ and $\chi_-$ priors with almost zero support at extreme values. The \emph{Flatmag} is singled out by the strong peak at $\chi_{+,-}=0$. As we will discuss in Sec.~\ref{results}, these non-trivial shapes of the priors can strongly affect the resulting posterior distributions in some cases.

%As we will discuss in section \ref{results}, the chirp mass is always very well constrained from observations (within $10^{-4}\%$). To extract the contribution of the prior to the posterior distribution under this constraint, in Fig. \ref{comp_priors} we display the \emph{Flatphys}, \emph{Flatmag} and \emph{Flatsampl} prior distributions for $\eta$, $\chi_+$ and $\chi_-$ obtained restraining the chirp mass to be within an interval. This interval is much broader than the typical measurement error so this would have no consequence on the posterior. The remarkable features of our fiducial prior, the \emph{Flatphys} prior, are the peaking of the \emph{Flatphys} prior at $\eta=0.25$ and $\eta=0$ and the bell shape for the $\chi_+$ and $\chi_-$ priors with almost zero support at extremal values. The \emph{Flatmag} is singled out by the strong peaking at $\chi_{+,-}=0$. As we will discuss in section \ref{results}, these features can strongly impact the PE.
%

\subsection{LISA response}

We briefly review how the LISA response is computed and refer to \cite{Marsat:2020rtl} for a more extensive discussion. We recall that LISA is composed of three spacecraft linked by lasers across arms of length $L=2.5 \ {\rm Gm}$.
The TDI observables $A$, $E$, and $T$ are time-delayed linear combinations of the single-link observables $y_{slr}$ that measure the laser frequency shift due to an incoming GW across link $l$ between spacecrafts $s$ and $r$. We consider only the dominant 22 mode of the waveform, and following~\cite{Marsat:2020rtl} we exploit a mode symmetry (for nonprecessing systems) between $h_{22}$ and $h_{2,-2}$ to write the signal in terms of $h_{22}$ only. The single-link observables can then be written with a transfer function: 
\begin{equation}
	\tilde{y}_{slr}=\mathcal{T}^{22}_{slr}\tilde{h}_{22} \,.
\end{equation}
Denoting the amplitude and phase of the mode 22 as $\tilde{h}_{22}=A_{22}(f)e^{-i\Psi_{22}(f)}$, working at leading order in the separation of timescales in the formalism of~\cite{Marsat:2018oam} the transfer functions are given by (we set $c=1$):
\begin{align}
	\mathcal{T}^{22}_{slr}&=G^{22}_{slr}(f,t_{f}^{22}) \\
	G^{22}_{slr}(f,t)&=\frac{i\pi fL}{2}{\rm sinc}\left[ \pi f L(1- {\bf k}\cdot {\bf n}_l) \right ] \nonumber  \\
& \cdot \exp \left [i\pi f \left(L+ {\bf k} \cdot ({\bf p}^L_{r}+{\bf p}^L_{s}) \right) \right ] \nonumber \\
& \cdot \exp(2i\pi f {\bf k} \cdot {\bf p}_0) \; {\bf n}_l \cdot {\bf P}_{22} \cdot {\bf n}_l   \label{kernel} \\
	t_{f}^{22}&=-\frac{1}{2\pi}\frac{{\rm d}\Psi_{22}}{{\rm d}f}, \label{eq:tf}
 \end{align}
 where ${\bf k}$ is the unit GW propagation vector, ${\bf n}_l(t)$ is the link unit vector pointing from the spacecraft $s$ to $r$, ${\bf p}_0(t)$ is the position vector of the center of the LISA constellation in the SSB frame, ${\bf p}^L_{r}(t)$ is the position of spacecraft $r$ measured from the center of the LISA constellation, ${\bf P}_{22}$ is the polarization tensor defined in \cite{Marsat:2020rtl} and we adopt the convention ${\rm sinc}(x)=\sin(x)/x$. We dropped the $t$ dependence in Eq.~\eqref{kernel} for more clarity. The global factor $\exp(2i\pi f {\bf k} \cdot {\bf p}_0)$ is the Doppler modulation in GW phase and the ${\bf n}_l \cdot {\bf P}_{22} \cdot {\bf n}_l$ term is the projection of the GW tensor on the interferometer axes, which is associated with the antenna pattern function. Note that both the Doppler modulation in phase and the antenna pattern are time dependent due to LISA's motion. Moreover, they depend on the sky position of the source, so that the annual variation in the phase and amplitude allows us to localize the source.

 \begin{figure*}
\centering
 \includegraphics[width=\textwidth]{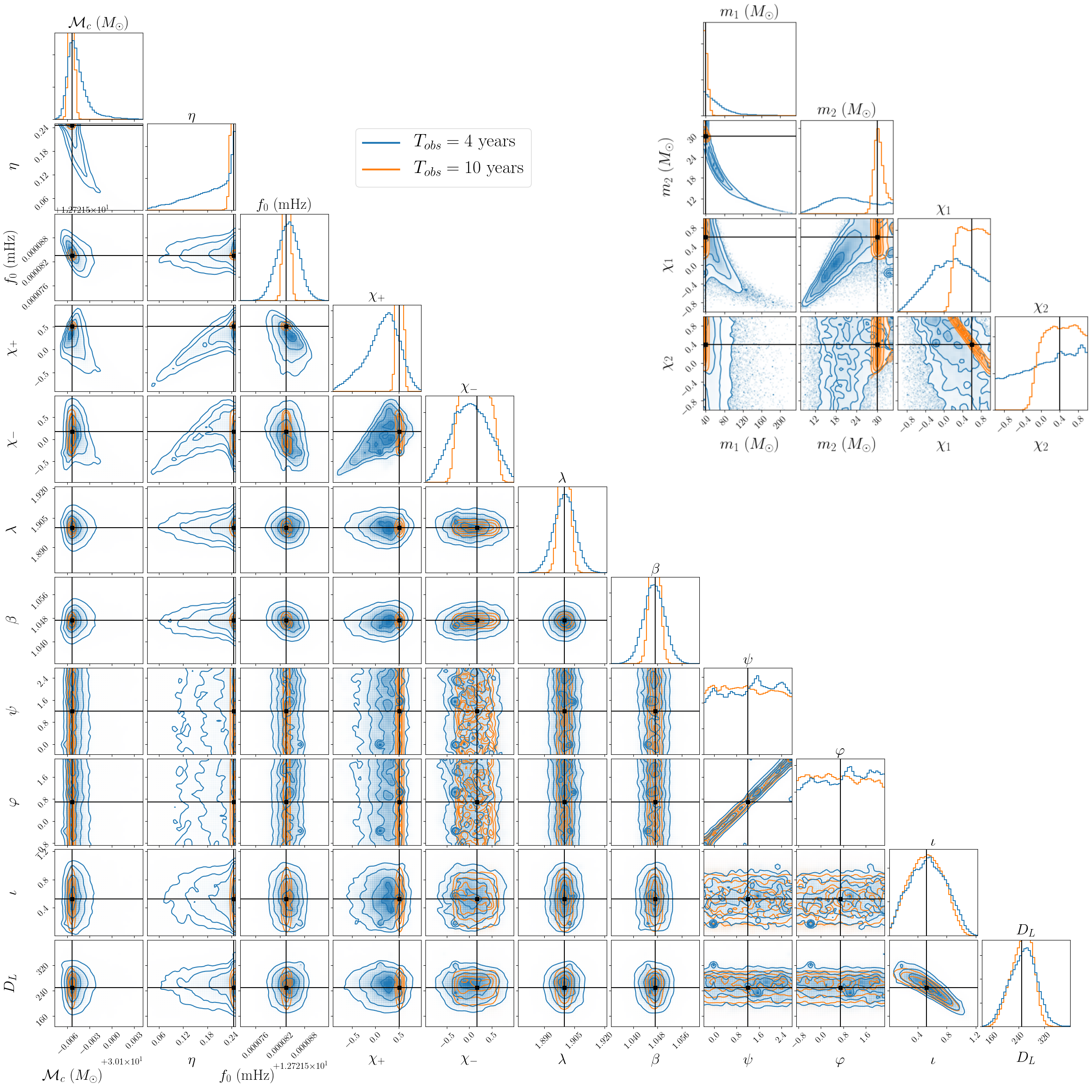}\\
 \centering
 \caption{Inferred parameter distribution for the \emph{Fiducial} system, both in the $T_{\rm obs}= 4\mathrm{yr}$ case (blue) and the $T_{\rm obs}=10\mathrm{yr}$ case (orange). The true parameters are indicated by black lines and squares. Masses are in the detector frame.}\label{comp_tobs}
\end{figure*}

All our results are obtained using the full LISA response, but we also assess the impact of using the long-wavelength approximation, a simplified version of the LISA response \cite{Cutler:1997ta}. In this approximation, LISA is somewhat similar to two LIGO/Virgo-type detectors rotated one with respect to the other by $\pi/4$, and with angles of $\pi/3$ between the arms. It is obtained by taking the $2\pi fL \ll 1$ limit in the LISA response so that:
\begin{align}
 G^{22}_{slr}(f,t) &= \frac{i\pi fL}{2} \exp(2i\pi f {\bf k} \cdot {\bf p_0}){\bf n}_l \cdot {\bf P}_{22} \cdot {\bf n}_l.  \label{kernel_lw}
\end{align}
The ${\rm sinc}$ function appearing in Eq.~\eqref{kernel} leads to a damping of the signal amplitude at high frequencies. However, in the long-wavelength approximation, it is replaced by 1, leading to unrealistically high SNRs. To compensate for this, inspired by the computation of the sky-averaged sensitivity \cite{Cornish:2018dyw}, we introduce a degradation function that multiplies the GW amplitude:
\begin{equation}
R(f)=\frac{1}{1+0.6(2\pi fL)^2}. \label{degrad_highf}
\end{equation}

To explore the validity of this approximation for SBHBs, we will compare the PE for the \emph{Fiducial}, \emph{Polar} and \emph{Equatorial} systems using the full response and the long-wavelength approximation labeled \emph{Full} and \emph{LW} respectively. We will only use the leading order in the separation of timescales in the framework of~\cite{Marsat:2018oam}, keeping in mind that corrections could be needed in general, in particular for almost-monochromatic signals.

\section{Parameter estimation of SBHBs}\label{results}

In order to test the performance of our MHMCMC sampler we compared it to our well tested parallel tempering MCMC code \texttt{PTMCMC} \footnote{https://github.com/JohnGBaker/ptmcmc}. The similarity of two distributions $p_1$ and $p_2$ can be quantified by computing their Kullback-Leibler (KL) divergence \cite{kullback1951}:
\begin{equation}
 D_{KL}=\sum_{\theta}p_1(\theta)\log \left ( \frac{p_1(\theta)}{p_2(\theta)} \right ).
\end{equation}
The KL divergence is zero if two distributions are identical.
%The closer it is from 0, the more $p_2$ is similar to the target distribution $p_1$.
We computed $D_{KL}$ for the marginalized distributions of each parameter obtained with two samplers using the \emph{Flatsampl} prior, and assuming four and ten years of observation. Apart from the polarization and the initial phase, all divergences were below $0.1$ for four years of observation and below $0.01$ for ten years of observation, showing a very good agreement between samplers. For $\psi$ and $\varphi$, less well determined in general, we get slightly higher values (up to $\simeq$ 0.6) but still showing a good agreement. The results presented in this paper were obtained with our MHMCMC code and, unless otherwise specified, we use the \emph{Flatphys} prior and the \emph{Full} response. In our discussion, we use redshifted masses (rather than source frame) because they are directly inferred from the observed data.
%refer to redshifted masses rather than source frame masses because those are the relevant quantities to analyse the structure of PE.
We give the full ``corner plot'' \cite{corner} for our fiducial system, comparing results for the two observation times in Fig.~\ref{comp_tobs}, this plot shows pair-wise correlation between parameters and the fully marginalized posterior for each parameter.
The inset on the right top of the figure shows posterior distributions for ($m_1,m_2,\chi_1,\chi_2$).
%All corner plots in this work were made with the {\bf corner} package \cite{corner}.

It would be difficult to represent the posterior distributions for all possible variations (deviations from the \emph{Fiducial}) discussed above. Instead, we will summarize our results by underlining qualitative differences whenever we observe them and show comparative corner plots only when necessary. We start by discussing the structure of correlation between intrinsic parameters, move to extrinsic parameters, then compare the full Bayesian analysis with predictions from the Fisher matrix, and finally show the effect of the \emph{LW} approximation to the response.
%
%The most straightforward way to verify how do correlations between parameters and the accuracy on each parameter change across the parameter space would be to plot distributions on top of each other. However, for sake of clarity we can not do this for all combinations we considered. Thus we will focus on discussing the main changes and show comparative corner plots only when necessary.
%We start by discussing the structure of correlation between intrinsic parameters and then move to extrinsic parameters. Following this we provide a comparison with Fisher matrices and then discuss how does the PE when using the low frequency approximation for LISA response.

\begin{figure}[!ht]
\centering
 \includegraphics[width=0.45\textwidth]{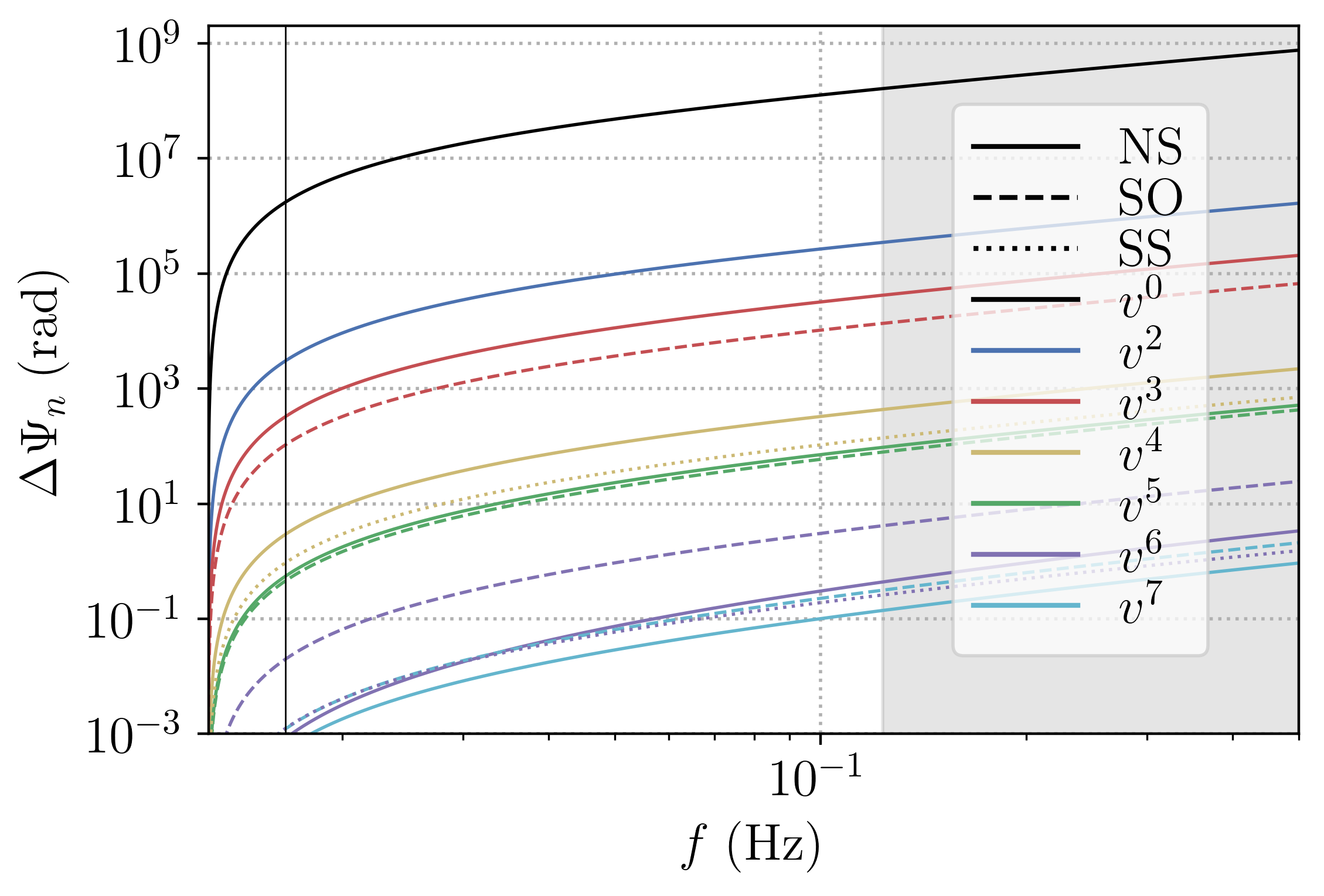}\\
 \centering
 \caption{Individual PN phase contributions $\Delta \Psi_{n}$ for the \emph{Fiducial} system. The linestyle indicates the nature of the term, nonspinning (NS), spin-orbit (SO) or spin-spin (SS), while the color indicates the PN order. Note that these contributions are individually aligned at $f_{0}$, as explained in the text, and that interpreting the magnitude of these terms is not easy due to the alignment freedom. The vertical line shows $f_{4\mathrm{yr}}$, and the greyed area shows the frequency range contributing less than 1 in $\mathrm{SNR}^{2}$.}\label{fig:phiPN}
\end{figure}

\begin{figure*}[!ht]
\centering
 \includegraphics[width=0.75\textwidth]{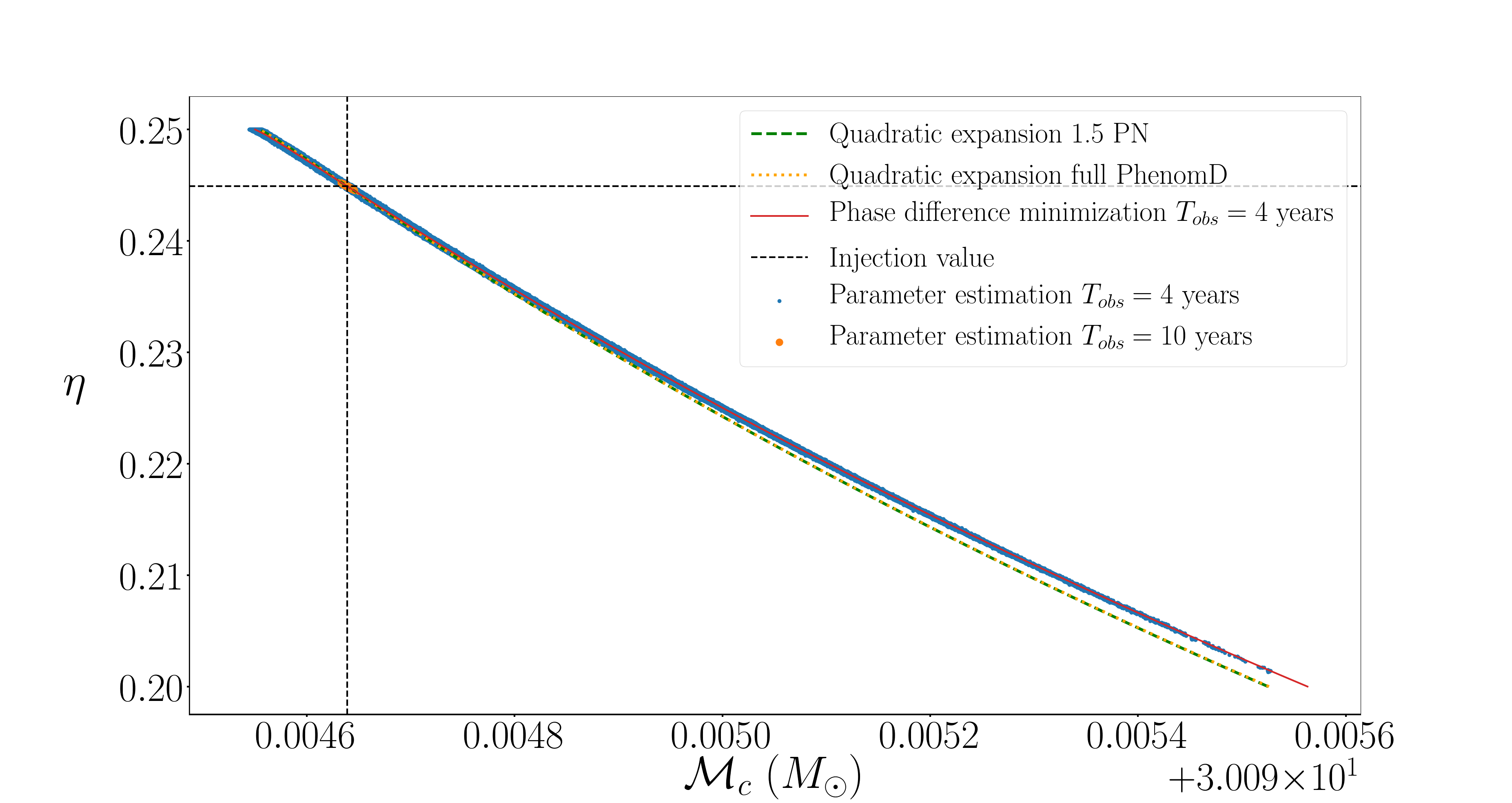}\\
 \centering
 \caption{Analysis of the degeneracy between $\chirpm$ and $\eta$. The blue (orange) dots were obtained running a PE on the \emph{Fiducial} system in the $T_{\rm obs}=4\mathrm{yr}$ ($T_{\rm obs}=10\mathrm{yr}$) case allowing only $\chirpm$ and $\eta$ to vary. The injection point is indicated by the black dashed lines. The orange dotted and the green dashed curves are given by \eqref{eq_degen} using the full PhenomD phase and the 1.5 PN truncation of the phase respectively. The red solid line was obtained by minimizing the phase difference between the injected signal and templates over the whole frequency range spanned over four years of observation.}\label{degen}
\end{figure*}

\subsection{Intrinsic parameters}\label{res_int}

One of the main features appearing in Fig.~\ref{comp_tobs} is the strong correlation between intrinsic parameters, in particular the one between $\chirpm$ and $\eta$ which is especially pronounced for four years of observation. The main reason for this degeneracy is the limited evolution of the GW frequency: in four years of observation the \emph{Fiducial} system spans a very narrow range from $f_0=12.7 \ {\rm mHz}$ to $f_{4 \mathrm{yr}}=16.5 \ {\rm mHz}$.

First, it will be instructive to consider the magnitude of the different PN orders appearing in the phasing (see~\cite{Blanchet:2002av} for a review). We can write formally
\begin{equation}
	\Psi (f) = \frac{3}{128 \eta v^{5}} \sum_{i} a_{i} v^{i} \,,
\end{equation}
where $v = (\pi M f)^{1/3}$ (with $M=m_1+m_2=\chirpm \eta^{-3/5}$) and where the $a_{i}$ are PN coefficients (we scaled out the leading term, so that $a_{0}=1$) that depend on the mass ratio and on the spins and can be separated between nonspinning terms (NS), spin-orbit terms (SO), and spin-square terms (SS). It was argued in~\cite{Mangiagli:2018kpu} that most SBHBs would require terms up to the 2 PN order. In Fig.~\ref{fig:phiPN}, we show the magnitude of the known PN terms in the phasing for our \emph{Fiducial} system. In general, the magnitude of phase contributions is delicate to interpret because of the alignment freedom, as some of the phasing error can typically be absorbed in a time and phase shift. In Fig.~\ref{fig:phiPN} we align contributions individually at $f_{0}$ with a zero phase and zero time according to~\eqref{eq:tf}. We see that, for $T_{\rm obs} = 4\mathrm{yr}$, PN orders beyond 1.5 PN appear negligible due to the limited chirping in frequency, while more terms become relevant for $T_{\rm obs} = 10\mathrm{yr}$ where much higher frequencies are reached. We also gray out the area ($f>123 \ {\rm mHz}$) beyond which the signal contributes less than 1 in $\mathrm{SNR}^{2}$, which we take as a somewhat conventional limit to indicate that ignoring the signal beyond this point would not affect the log-likelihood~\eqref{loglike} and therefore the PE.

In order to provide an explanation for the strong correlation between chirp mass and symmetric mass ratio, we consider a simplified problem by reducing the dimensionality: we fix $f_0$, $\chi_+$, $\chi_-$ and all extrinsic parameters to the ``true'' values and investigate the correlation between the chirp mass and the symmetric mass ratio for the \emph{Fiducial} system. Keeping these parameters fixed will collapse some of the degeneracies seen in the full analysis, but this exercise will serve as an illustration of the differences between a nonchirping and chirping system.

Since for $T_{\rm obs} = 4\mathrm{yr}$ the GW frequency changes little from $f_0$ to $f_{4 \mathrm{yr}}$, we can Taylor expand the phase around $f_0$:
%perform a PE on $\chirpm$ and $\eta$ for the \emph{Fiducial} system.
\begin{equation}
 \Psi(f) \simeq \Psi(f_0)+ \left . \frac{{\rm d} \Psi}{{\rm d}f} \right |_{f_0}(f-f_0)+\frac{1}{2} \left . \frac{{\rm d^2} \Psi}{{\rm d}f^2} \right |_{f_0}(f-f_0)^2.
\end{equation}

We consider the inner product between the data $d=A_d(f) e^{-i\Psi_d(f)}$ and the template $h=A_h(f) e^{-i\Psi_h(f)}$. From our convention, the initial phase at $f_{0}$ is the same, $\Psi_d(f_0) = \Psi_h(f_0)$. The initial time is zero at $f_{0}$, so the stationary phase approximation Eq.~\eqref{eq:tf} gives: $ \frac{{\rm d} \Psi}{{\rm d}f} |_{f_0}=0$. The inner product becomes:
\begin{align}
 (d|h)& = 4\mathrm{Re}\int_{f_0}^{f_{4 \mathrm{yr}}} df \frac{A_d(f)A_h(f) e^{i(\Psi_d(f)-\Psi_h(f))} }{S_n(f)} {\rm d}f \nonumber  \\
 & \simeq A_d(f_0) A_h(f_0) \; 4 \mathrm{Re} \left[ \int_{f_0}^{f_{4 \mathrm{yr}}} \frac{df}{S_n(f)} e^{i(\frac{{\rm d^2} \Psi_d}{{\rm d}f^2} |_{f_0} - \frac{{\rm d^2} \Psi_h}{{\rm d}f^2} |_{f_0})\frac{(f-f_0)^2}{2}} \right],
\end{align}
where we used the fact that the amplitude is a slowly varying function of the frequency. The overlap is maximized when the template is in phase with the data, making the integrand nonoscillating. In our quadratic approximation to the dephasing, this defines a curve in the ($\chirpm$, $\eta$) plane according to
\begin{equation}
\left . \frac{{\rm d^2} \Psi}{{\rm d}f^2}  \right |_{f_0} = \left . \frac{{\rm d^2} \Psi (\mathcal{M}_{c,0},\eta_{0})}{{\rm d}f^2}   \right |_{f_0} \label{eq_degen}.
\end{equation}

In Fig.~\ref{degen} we display in blue (orange) dots points from the sampling in the $(\chirpm,\eta)$ plane in the $T_{\rm obs}=4\mathrm{yr}$ ($T_{\rm obs}=10\mathrm{yr}$) case and overplot (in orange dotted line) the curve obtained by solving \eqref{eq_degen}. The true (injection) value is indicated by black dashed lines. The curve closely follows the shape obtained from PE in the $T_{\rm obs}=4\mathrm{yr}$ case. The green dashed line is obtained by solving \eqref{eq_degen} truncating the phase to 1.5 PN order. We verified that adding higher PN terms does not produce any noticeable changes, which is in a good agreement with \cite{Mangiagli:2018kpu} and Fig.~\ref{fig:phiPN}.

\begin{figure}[!ht]
\centering
 \includegraphics[width=0.45\textwidth]{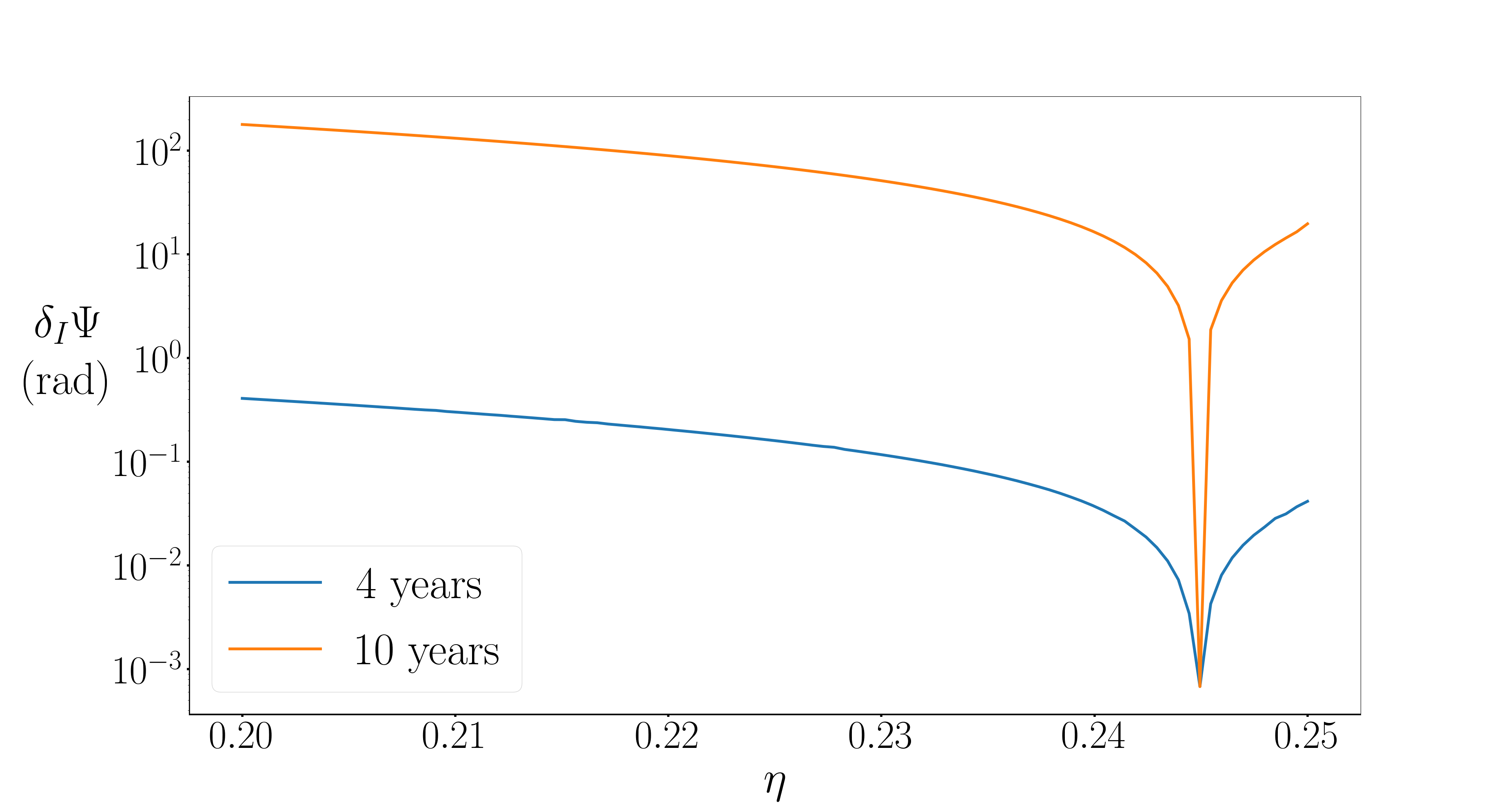}\\
 \centering
 \caption{Value of $\delta_{I} \Psi$ along the curve in the $(\chirpm,\eta)$ plane that minimizes it for $I=[f_0,f_{4 \ {\rm years}}]$ (blue) and $I=[f_0,f_{{\rm max,LISA}}]$ (orange). When LISA observes the system at low frequencies, the phase difference can be kept small over an extended region far from the injection. When LISA observes the chirp of the system, the phase difference becomes very large immediately at the vicinity of the injection point, reducing the extent of the degeneracy between $\chirpm$ and $\eta$.}\label{diff_psi}
\end{figure}

% Up to 1.5 PN, the second derivative is given by:
%
%%Analytically this curve is given by the equation
%
%\begin{align}
% \left . \frac{{\rm d^2} \Psi}{{\rm d}f^2} \right |_{f_0}&=\frac{5}{378} ( 40 (\pi \chirpm f_0)^{-5/3} +18\phi_2\eta^{-2/5}(\pi \chirpm f_0)^{-1}\\
% &+10\phi_3\eta^{-3/5}(\pi \chirpm f_0)^{-2/3}),
%\end{align}
%
%where:
%
%\begin{align}
% \phi_2&=\frac{3715}{756}+\frac{55 \eta}{9} \\
% \phi_3&=-16 \pi+\frac{94}{3}\chi_++ \frac{19}{3} \frac{q-1}{q+1} \chi_-.
%\end{align}
We can even better reproduce the degeneracy by minimizing the phase difference between injection and template over the whole frequency range spanned by the injected signal. More specifically, defining:
\begin{equation}
 \delta_{I} \Psi(\chirpm,\eta) ={\rm max}_I|\Psi(\mathcal{M}_{c,0},\eta_{0})(f)-\Psi(\chirpm,\eta)(f)|,
\end{equation}
for each value of $\chirpm$ we find $\eta$ such that $\delta_{I} \Psi$ is minimized. Note that all parameters are kept fixed in the dephasing measure we use here, in particular there is no optimization over a constant phase or time shift. The subscript $I$ stands for the frequency interval and we plot this curve for $I= [f_0,f_{4 \ {\rm years}}]$ in Fig.~\ref{degen}. One can see that we almost perfectly reproduce the shape of the correlation between the chirp mass and the symmetric mass ratio in the $T_{\rm obs}=4\mathrm{yr}$ case.
In the $T_{\rm obs}= 10\mathrm{yr}$ case, the system evolves until it leaves the band so it spans a broader frequency range. In Fig.~\ref{diff_psi} we show the value of the minimised $\delta_I \Psi$ for $I= [f_0,f_{4 {\rm years}}]$ and for $I=[f_0,f_{\rm max}^{\rm LISA}]$ with $f_{\rm max}^{\rm LISA}=0.5 \ {\rm Hz}$ taken at the conventional end of the LISA frequency band. In practice, in the latter case the maximal dephasing occurs typically around $\sim 0.1\ {\rm Hz}$. For the observation span of four years, we can find $\delta_I \Psi$ to be quite small ($<0.5 \  {\rm rad}$) over a large range of $\eta$. As the bandwidth of the signal becomes broader, we cannot efficiently compensate for a change in the chirp mass by varying $\eta$, which results in a significant reduction of the degeneracy and a great improvement in measuring those two parameters, as seen by the narrower region covered by the orange dots on Fig.~\ref{degen}.

%
%, it becomes impossible to maintain $\delta \Psi$ small far from the true point. In this case the measurement of $\chirpm$ and $\eta$ is greatly improved.
%
%To see how this impacts the phase difference, we show on Fig. \ref{diff_psi} the value of $\delta_I \Psi$ along the curve minimising it both for $I= [f_0,f_{4 years}]$ and for $I=[f_0,f_{max,LISA}]$.
%For an observation time of $4 \ {\rm years}$, we can keep $\delta_I \Psi$ small ($<0.5 \  {\rm rad}$) in an extended region far from the true point.
%As the frequency range becomes broader, it becomes impossible to maintain $\delta \Psi$ small far from the true point. In this case the measurement of $\chirpm$ and $\eta$ is greatly improved.

\begin{figure}[!ht]
\centering
 \includegraphics[width=\columnwidth]{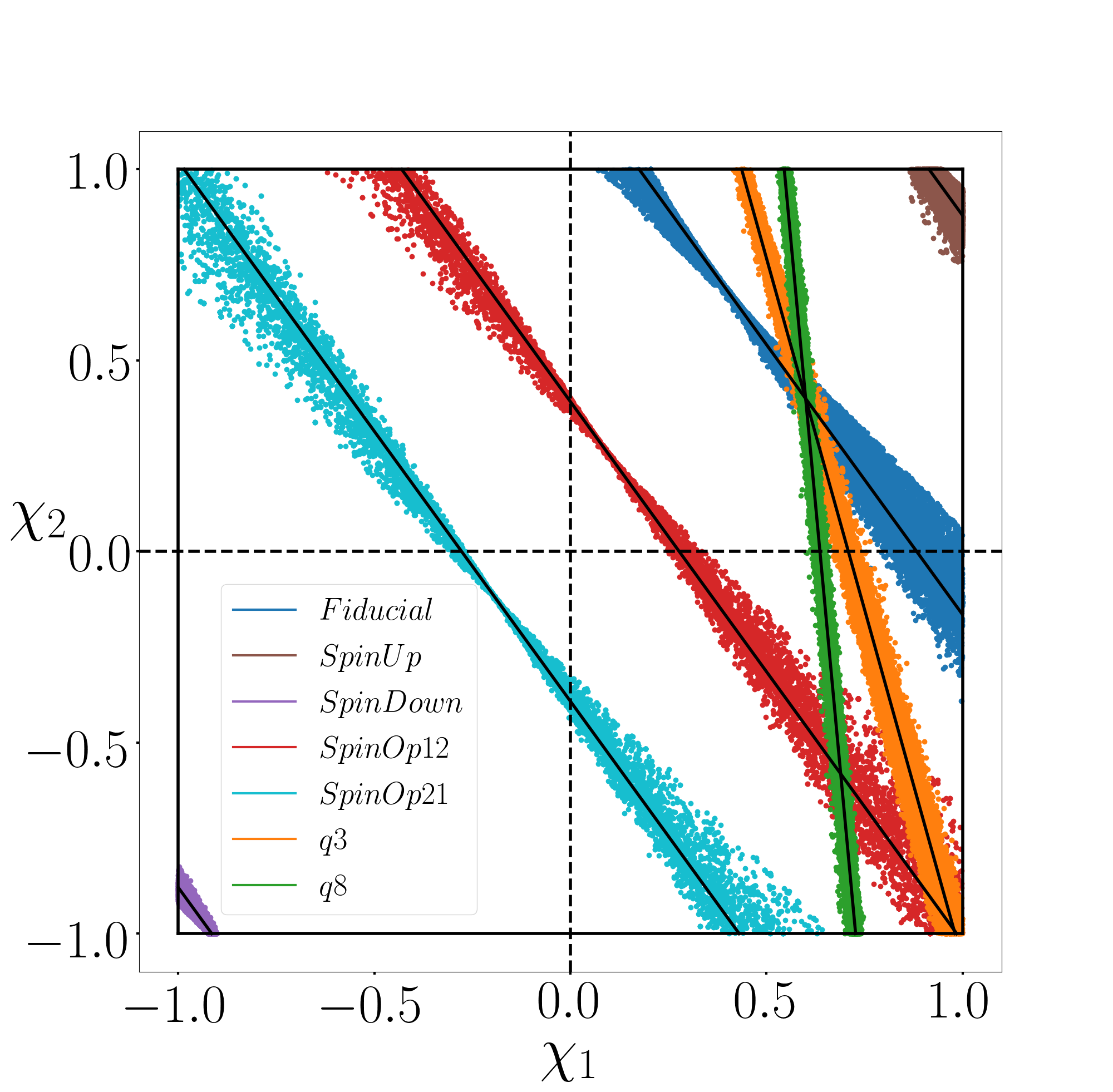}
 \centering
 \caption{Samples of $\chi_1$ and $\chi_2$ obtained for different systems (defined in Sec.~\ref{setups}) in the $T_{\rm obs}=10\mathrm{yr}$ case. The black solid lines indicates the boundaries of the physically allowed region $-1 \leq \chi_{1,2} \leq 1$ and the $\chi_{\rm PN}=\chi_{\rm PN,0}$ lines. The samples follow the $\chi_{\rm PN}=\chi_{\rm PN,0}$ lines, showing that this is the specific combination of spins that can be measured. The orthogonal combination of spins is constrained only due to the boundaries of the physically allowed region. Due to the orientation of the $\chi_{\rm PN}={\rm const}$ lines, $\chi_1$ is better constrained than $\chi_2$. High values of spins with same (opposite) sign are the better (worse) constrained.}\label{chischia}
\end{figure}

We now come back to the full Bayesian analysis and consider the estimation of the BH spins. Following \cite{Poisson:1995ef,Khan:2015jqa} we introduce the 1.5 PN spin combination:
\begin{align}
 \chi_{\rm PN}&=\frac{1}{113} \left ( 94\chi_++19 \frac{q-1}{q+1}\chi_- \right ) \\
 &=\frac{\eta}{113} \left ( (113q+75)\chi_1+(\frac{113}{q}+75)\chi_2 \right ).\label{chipn}
\end{align}
This term defines how the spins enter the GW phase at the leading (1.5 PN) order~\cite{Blanchet:2002av} and, therefore, should be
the most precisely measured spin combination.
 %so that the 1.5 PN coefficient simply reads $\varphi_3=-16 \pi + \frac{113}{3} \chi_{\rm PN}$.
%  It is in fact this particular combination that can be measured when observing SBHBs at higher frequencies.
We found this to be indeed the case. As an illustration, we plot samples obtained for the \emph{q3}, \emph{q8}, \emph{SpinUp}, \emph{SpinDown}, \emph{SpinOp12} and \emph{SpinOp21} systems in the $T_{\rm obs}=10\mathrm{yr}$ case in Fig.~\ref{chischia}. The points are the samples obtained in PE analysis and the lines show $\chi_{\rm PN}=\chi_{\rm PN,0}$ (fixing the mass ratio to its ``true'' (injection) value) for all those systems in the ($\chi_1$, $\chi_2$) plane.
%Samples follow the constant $\chi_{\rm PN}$ lines and are limited by the prior boundaries \st{bounds on what? spins?}.
In all these cases $\chi_{\rm PN}$ is extremely well measured, within $10^{-2}$, but the combination of spins orthogonal to $\chi_{\rm PN}$ is constrained only by the prior boundaries.

For slowly evolving binaries, only terms up to 1.5 PN in the GW phase are found to be relevant. At this order we expect a strong correlation between the 1.5 PN spin combination and the symmetric mass ratio: any change in $\chi_{\rm PN}$ can be efficiently compensated by a change in $\eta$ such that the 1.5 PN term $(-16 \pi + {113}{3}\chi_{\rm PN})\eta^{-3/5}$ is kept
(almost) constant. We have verified this by plotting the curve $(-16 \pi + {113}{3}\chi_{\rm PN})\eta^{-3/5}=\rm{const}$ on top of the samples obtained for the \emph{Fiducial} system and reproducing the shape formed by the posterior samples. Thus, we obtained and explained the three-way correlation between chirp mass, mass ratio and spins for the mildly relativistic systems spanning a narrow frequency band during the observation time. The increase in the observation time allows further chirping of the system, making the contribution of the 1 and 1.5 PN corrections in the phasing significant, thus breaking strong correlations between intrinsic parameters; however, the effect of higher-order PN terms is weak, consistently with~\cite{Mangiagli:2018kpu} and Fig.~\ref{fig:phiPN}, which leads to only the 1.5 PN spin combination being measured. This study also suggests that $\chi_{\rm PN}$, being the most relevant mass-weighted spin combination for PE, should be used as sampling parameter. The component of $\chi_{\rm PN}$ along $\chi_+$ is always much larger than the one along $\chi_-$ (at least by a factor $\frac{94}{19}\simeq 5$), so we find that $\chi_+$ is also measured reasonably well. The effective spin $\chi_+$ is frequently used in the GW literature and has a clear astrophysical interpretation, as opposed to the 1.5 PN spin combination; therefore, we will alternate between $\chi_{\rm PN}$ and $\chi_+$ in our next discussions.

\begin{figure}[!ht]
\centering
 \includegraphics[width=0.5\textwidth]{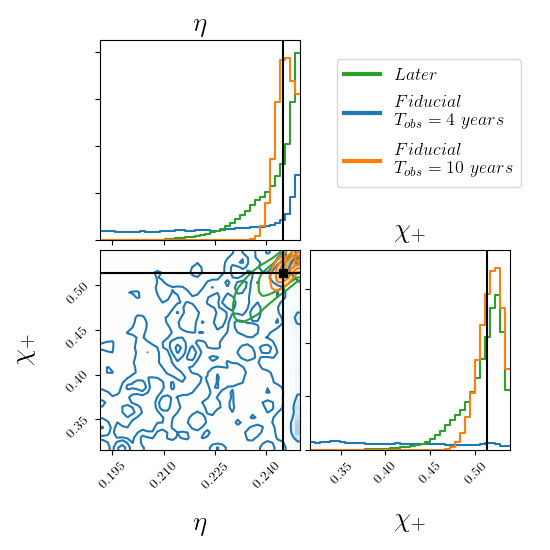}\\
 \centering
 \caption{Distribution of $\eta$ and $\chi_+$ for the \emph{Later} system ($t_{c} = 2\mathrm{yr}$) and the \emph{Fiducial} system ($t_{c} = 8\mathrm{yr}$) for both observation times ($T_{\rm obs} = 4\mathrm{yr}$ and $T_{\rm obs} = 10\mathrm{yr}$). Since we observe the \emph{Later} system chirping, the determination of $\eta$ and $\chi_+$ is much better than for the \emph{Fiducial} system in the $T_{\rm obs}=4\mathrm{yr}$ case. But because of its low SNR ($\mathrm{SNR}=11.8$), the posterior distribution still peaks at $\eta=0.25$, as an effect of the prior. This is on contrast to the \emph{Fiducial} system in the $T_{\rm obs}=10\mathrm{yr}$ case ($\mathrm{SNR}=21.1$) which peaks at the injected value indicated by black lines and squares.}\label{etas}
\end{figure}

\begin{table*}
 \begin{center}
   \begin{tabular}{c *{17}{c|}}
   \cline{3-17}

  & &  \multicolumn{5}{|c|}{\emph{Fiducial}} & \multicolumn{5}{|c|}{\emph{Earlier}} & \multicolumn{5}{|c|}{\emph{Later}} \\

  \cline{3-17}

   &  & $\chirpm$ & $\eta$ & $\chi_+$ & $\chi_-$ & $\chi_{\rm PN}$ & $\chirpm$ & $\eta$ & $\chi_+$ & $\chi_-$ & $\chi_{\rm PN}$ & $\chirpm$ & $\eta$ & $\chi_+$ & $\chi_-$ & $\chi_{\rm PN}$  \\

   \hline

  \multicolumn{1}{|c|}{\multirow{2}{*}{\emph{Flatphys}}} & \multicolumn{1}{|c|}{$T_{\rm obs}=4\mathrm{yr}$} & $3.6$ & $0.4$ & $0.2$ & $0.1$ & $0.3$ & $2.7$ &$0.04$ & $0.04$ & $0.03$ &  $0.04$ &$6.1$ & $1.7$ & $3.1$ & $0.4$ & $3.6$ \\

   \cline{2-17}

   \multicolumn{1}{|c|}{} & \multicolumn{1}{|c|}{$T_{\rm obs}=10\mathrm{yr}$} & $7.6$ & $2.5$ & $3.7$ & $0.5$ & $4.3$ & $4.5$ & $0.7$ & $0.5$ & $0.2$ & $0.6$ &/ & / & / & / & /\\

   \hline\hline

    \multicolumn{1}{|c|}{\multirow{2}{*}{\emph{Flatmag}}} & \multicolumn{1}{|c|}{$T_{\rm obs}=4\mathrm{yr}$} & $3.4$ & $0.6$ & $0.07$ & $0.04$ & $0.08$ & / & / & / & / & /
  & / & / & /  & / & / \\

   \cline{2-17}

   \multicolumn{1}{|c|}{} & \multicolumn{1}{|c|}{$T_{\rm obs}=10\mathrm{yr}$} & $7.5$ & $2.5$ & $4.4$ & $0.4$ & $4.8$ & / & / & / & / & / & / & / & / & / & /\\

   \hline\hline

    \multicolumn{1}{|c|}{\multirow{2}{*}{\emph{Flatsampl}}} & \multicolumn{1}{|c|}{$T_{\rm obs}=4\mathrm{yr}$} & $3.7$ & $0.4$ & $0.3$ & $0.2$ & $0.3$ & / & / & / & / & /
  & / & / & / & /  & / \\

   \cline{2-14}

   \multicolumn{1}{|c|}{} & \multicolumn{1}{|c|}{$T_{\rm obs}=10\mathrm{yr}$} & $7.3$ & $3.2$ & $3.7$ & $0.5$ & $4.4$ &/ & / & / & / & / & / & / & / & / & / \\

   \hline

   \end{tabular}
\end{center}
 \caption{Kullback-Leibler divergences between the marginalized posterior and prior distribution of the intrinsic parameters for different systems and choices of prior. When observing the system at low frequencies, only $\chirpm$ shows a sensible deviation from the prior. The likelihood is informative on $\eta$ and $\chi_+$ (and $\chi_{\rm PN}$) only for chirping systems. Different choices of prior give similar results.}\label{kl_test}
\end{table*}

In order to further quantify the dependence of PE on the frequency bandwidth spanned by the signal during the observation time, we consider the \emph{Earlier}, \emph{Fiducial} and \emph{Later} systems which differ in the initial frequency chosen so that the SBHBs merge in 20, 8 and 2 years respectively. We compute the KL divergence between the marginalized posterior and the marginalized prior for each intrinsic parameter, and report our findings in Table~\ref{kl_test}. The larger values of $D_{KL}$ indicate that knowledge has been gained from the GW observations as compared to the prior. The results show a strong dependence on the observation time (therefore on the frequency bandwidth), especially for spins, for which the $D_{KL}$ varies by an order of magnitude. For the \emph{Earlier} system we find that only the chirp mass measurement is truly informative. Note that the longer frequency evolution plays a bigger role than the SNR. For instance, \emph{Later} which leaves LISA after two years with $\mathrm{SNR}=11.8$ is more informative than \emph{Earlier} with $T_{\rm obs}=10\mathrm{yr}$ which has an $\mathrm{SNR}=17.2$. We repeated this analysis using the \emph{Flatmag} and \emph{Flatsampl} priors for the \emph{Fiducial} system. For all choices of prior, the KL divergences are similar, proving the $\eta$, $\chi_+$, $\chi_-$ distributions are prior dominated when observing slowly evolving systems. Notice that KL divergences for spins are slightly smaller when using the \emph{Flatmag} prior, meaning that the posterior is even more dominated by the prior. This is because the \emph{Flatmag} prior peaks strongly at $\chi_+=\chi_-=0$ as discussed in Sec.~\ref{priors}. Note that the values of $D_{KL}$ are always larger for $\chi_{\rm PN}$ than for the other spin combinations, reflecting the fact that it is the best measured spin combination. Still, for systems evolving through a narrow frequency interval, the $\chi_{\rm PN}$ distribution is also prior dominated. The effect of the prior is especially well seen for the \emph{Fiducial} system and $T_{\rm obs}= 4\mathrm{yr}$ in Fig.~\ref{etas}: the strong peak of the symmetric mass ratio at 0.25 is what we expect due to prior (see Sec.~\ref{priors}). The same peak is also observed for the \emph{Later} system (predominantly due to low SNR) but $\eta$ is much better constrained for this system, the likelihood is informative enough to reduce the width of the distribution, but not large enough to supress the prior. {\it Let us reiterate this important finding: for intrinsic parameters beyond the chirp mass, the chirping (extent of the frequency evolution) of the observed SBHB has stronger influence on PE than the SNR or observation time per se.}

\begin{figure*}
\centering
 \includegraphics[width=0.49\textwidth]{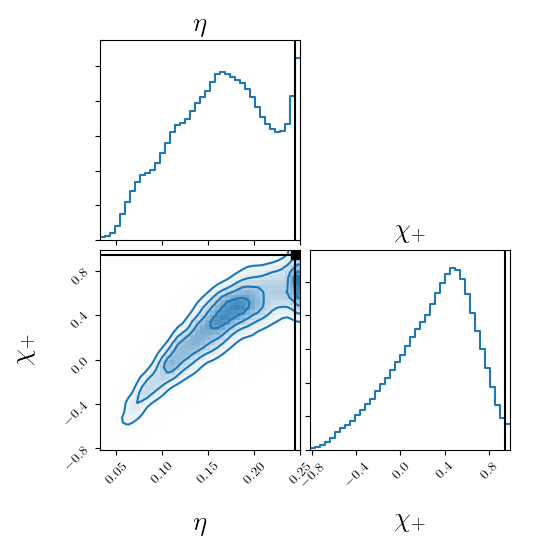}
 \includegraphics[width=0.49\textwidth]{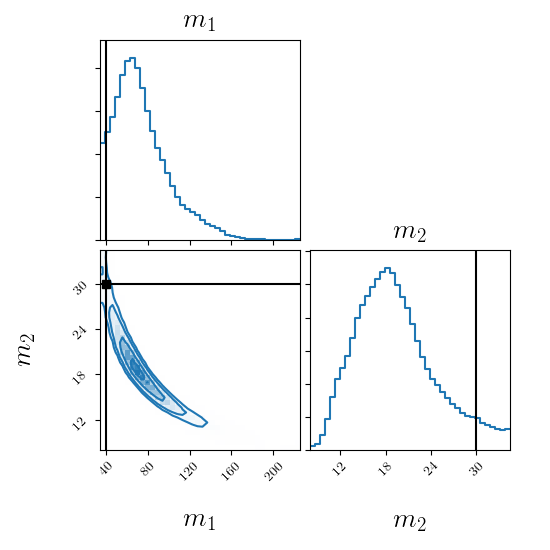}
 \centering
 \caption{The left panel shows the inferred distribution on $\eta$ and $\chi_+$ for the \emph{SpinUp} system. Because of a ``competition'' between the prior and the likelihood the distributions of $\eta$ and $\chi_-$ peak away from the true value indicated by the black lines and and the square. The $\chirpm$ distribution, not shown, is marginally affected. Because of the bias in $\eta$, the inferred distribution of masses is significantly biased. However with our definitions, the true value is within the $90 \%$ CI.}\label{degen_chis_eta}
\end{figure*}

\begin{table}
  \begin{center}
   \begin{tabular}{c|c|c|}

   \cline{2-3}

    & $T_{\rm obs}=4\mathrm{yr}$ & $T_{\rm obs}=10\mathrm{yr}$ \\

    \hline

    \multicolumn{1}{|c|}{$\chirpm/ \mathcal{M}_{c,0}$} & $1^{+1\times 10^{-4}}_{-4 \times 10^{-5}}$ & $1^{+2 \times  10^{-6}}_{-1 \times  10^{-6}}$ \\

    \hline

     \multicolumn{1}{|c|}{$\mathcal{M}_{c,s}/ \mathcal{M}_{c.s,0}$} & $0.99^{+0.01}_{-0.01}$ & $1.00^{+0.01}_{-0.01}$ \\

    \hline

    \multicolumn{1}{|c|}{$q$} & $2.6^{+4.7}_{-1.6}$ & $1.3^{+0.1}_{-0.3}$ \\

   \hline

   \multicolumn{1}{|c|}{$m_1 / m_{1,0}$} & $1.4^{+1.1}_{-0.6}$ & $0.99^{+0.04}_{-0.13}$ \\

   \hline

   \multicolumn{1}{|c|}{$m_2 / m_{2,0}$} & $0.7^{+0.4}_{-0.3}$ & $1.06^{+0.14}_{-0.04}$ \\

   \hline

    \multicolumn{1}{|c|}{$m_{1,s} / m_{1,s,0}$} & $1.5^{+1.2}_{-0.6}$ & $0.99^{+0.04}_{-0.13}$ \\

   \hline

   \multicolumn{1}{|c|}{$m_{2,s} / m_{2,s,0}$} & $0.7^{+0.5}_{-0.3}$ & $1.06^{+0.15}_{-0.04}$ \\

   \hline
   %\multicolumn{1}{|c|}{$f_0/ f_{0,0}$} & $1.0^{+3.2 10^{-7}}_{-3.4.10^{-7}}$ & $0.239$ \\
   %\hline

    \multicolumn{1}{|c|}{$\chi_+$} & $0.2^{+0.5}_{-0.7}$ & $0.52^{+0.01}_{-0.02}$ \\

   \hline

   \multicolumn{1}{|c|}{$\chi_-$} & $0.03^{+0.7}_{-0.6}$ & $0.1^{+0.4}_{-0.4}$ \\

    \hline

     \multicolumn{1}{|c|}{$\chi_{\rm PN}$} & $0.2^{+0.4}_{-0.7}$ & $0.433^{+0.008}_{-0.009}$ \\

    \hline

   \multicolumn{1}{|c|}{$\chi_1$} & $0.2^{+0.8}_{-0.6}$ & $0.6^{+0.4}_{-0.3}$ \\

   \hline

   \multicolumn{1}{|c|}{$\chi_2$} & $0.2^{+0.7}_{-1.0}$ & $0.4^{+0.6}_{-0.5}$ \\

   \hline

   \multicolumn{1}{|c|}{$\Delta t_c \ ({\rm s})$} & $10^4$ & $20$ \\

   \hline

   \multicolumn{1}{|c|}{$\Delta \Omega \ ({\rm deg}^2) $} & $0.18$ & $0.03 $  \\

   \hline

   \multicolumn{1}{|c|}{$D_L/D_{L,0}$} &  $1.1^{+0.2}_{-0.3}$ & $1.0^{+0.2}_{-0.2}$ \\

   \hline

    \multicolumn{1}{|c|}{$z$} &  $0.060^{+0.012}_{-0.014}$ & $0.055^{+0.009}_{-0.012}$ \\

   \hline

\end{tabular}
   \end{center}
    \caption{$90 \%$ CI on the parameters of the \emph{Fiducial} system whose parameters are given in Table \ref{params_fiducial} using the \emph{Flatphys} prior. For masses and distance, we give the relative errors. The redshifted chirp mass is extremely well determined for both mission durations but individual masses can be measured only if the mission is long enough and we can observe the system chirping. The measurement of the source frame chirp mass is worse, being dominated by the error on the distance measurement and therefore the redshift in~\eqref{rel_source_mass}. The error on individual masses is dominated by their intrinsic degeneracy. For chirping systems, we can also measure $\chi_{\rm PN}$ which translates into a good constraint on the effective spin $\chi_+$. The error on individual spins remains large for the chirping system, but we can start to constrain the spin of the primary BH (in our example, excluding negative values). As a consequence of the overall improvement in the determination of the intrinsic parameters, the inference of the time to coalescence improves drastically. The sky location (given by Eq.~\ref{eq_omega}) is very well determined for both mission durations, within the field of view of next generation electromagnetic instruments like Athena and SKA \citep{WinNT,2018CoSka..48..498M}. }\label{errors}
  \end{table}

We note that, although the frequency is slowly evolving, the signal is far from monochromatic unlike most of galactic binaries (e.g. double withe dwarf binaries). As an element of comparison, using the quadrupole formula to compute the frequency derivative at $f_0$, for the \emph{Earlier} system we find $\dot{f}_0=1.9 \times 10^{-11} \ {\rm Hz}^2$ which is four orders of magnitude higher than the fastest evolving galactic binaries \cite{Korol:2017qcx}. Thus, despite the strong correlation between intrinsic parameters, the chirp mass is always well measured, with a relative error of order $10^{-4}$ for the \emph{Earlier} system when observing for four years and below $10^{-6}$ for the chirping systems. The tight constraint on $\chirpm$ leads to the bananalike shape correlation between $m_1$ and $m_2$ seen on the top right part of Fig.~\ref{comp_tobs}. As a result, we can determine individual masses (within $20$--$30 \%$) only for chirping systems.

We give the $90\%$ CI for parameters of the \emph{Fiducial} system in Table \ref{errors}. Whenever the marginalized distribution of a given parameter is leaning against the upper (lower) boundary of the prior as for $m_1$ ($m_2$) we define the $90 \%$ CI as the value between the 0.1 and 1 quantile (0 and 0.9). Otherwise, in all other situations we define the $90 \%$ CI as the values between the 0.05 and 0.95 quantiles. In all cases we report the median as a point estimate.

\begin{figure*}
\centering
 \includegraphics[width=0.49\textwidth]{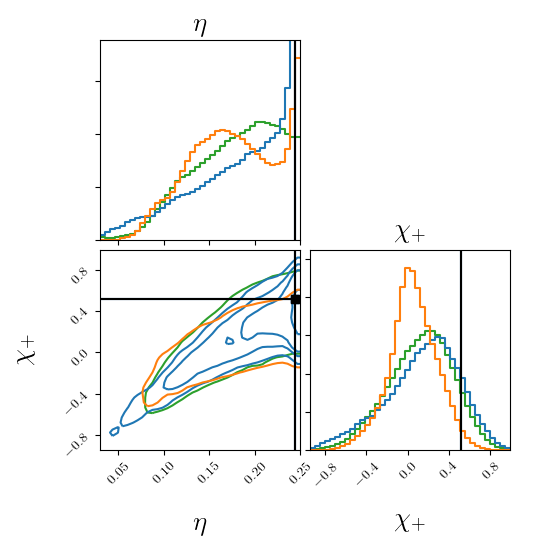}
 \includegraphics[width=0.49\textwidth]{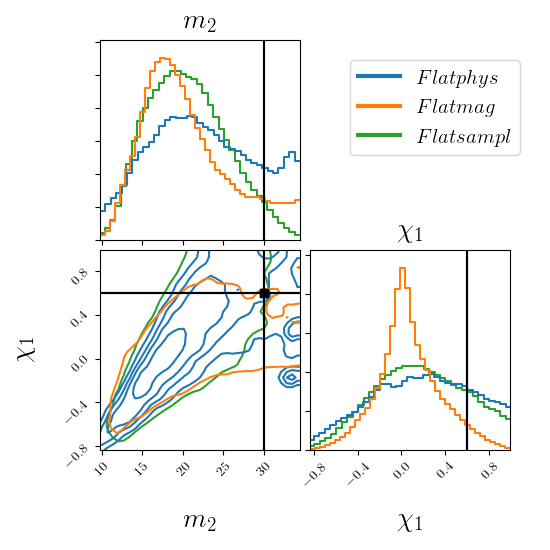}
 \centering
 \caption{The left (right) panel shows the inferred distribution on $\eta$ and $\chi_+$ ($m_2$, $\chi_1$) for the \emph{Fiducial} system using the \emph{Flatphys}, \emph{Flatmag} and \emph{Flatsampl} priors. Under the effect of the prior, the posterior distribution can be significantly shifted away from the true value indicated by black lines and squares.}\label{priors_post}
\end{figure*}

%The tight constraint on $\chirpm$ leads to the banana shape correlation between $m_1$ and $m_2$ seen on the top right part of Fig. \ref{comp_tobs}. Because two distinct combinations of masses are needed in order to determine individual masses, they can only be measured if we see the system chirping. In that case they are measured within $20-30 \%$. For non chirping systems the error on individual masses is above $100 \%$.
Systems with a higher mass ratio (\emph{q3} and \emph{q8}, keeping the chirp mass the same as for \emph{Fiducial}) give an error on the chirp mass similar to the \emph{Fiducial} system, but the mass ratio is better determined. This is because, when keeping the chirp mass fixed, the PN expansion of the GW phase features negative powers of $\eta$, notably in the 1 PN term. Moreover, what should matter is the derivative of the phase with respect to $\eta$ which contains 
only negative powers ($\eta^{-7/5}, \eta^{-2/5}$) which makes it more sensitive to the small mass ratio as compared to the equal-mass systems.
 %values of eta  This makes the overall dependency of the phasing on $\eta$ sharper for small mass ratios than for equal-mass systems. 
 %in the PN expansion, since higher mass ratios correspond to lower values of $\eta$, enhancing the contribution of $\eta$ to the phase.
For an observation time of four years, the uncertainty on individual masses is still of order of $100\%$, but for an observation time of ten years, it reaches below $10 \%$ and $1\%$ for the \emph{q3} and \emph{q8} system, respectively.

We now discuss the effect of priors on PE for high-spin systems. Consider \emph{SpinUp} system in the $T_{\rm obs}=4\mathrm{yr}$ case
 shown in Fig.~\ref{degen_chis_eta}. As discussed above, in this case we have the correlation between spin ($\chi_{\rm PN}$) mass ratio
 $\eta$ and the chirp mass. In the posterior we observe the interplay between the symmetric mass ratio and effective spin priors which push samples towards $\eta=0.25$ and $\chi_+=0$ and the likelihood which peaks at the true value of $\chi_+$ (0.95). This, together with the correlation between parameters, leads to the resulting posterior distribution which has double peak in $\eta$ and broad distribution for $\chi_+$ (the 2D histogram is more informative). The distribution (overall) is shifted away from the true values (well evident in the right panel of Fig.~\ref{degen_chis_eta}), though they are still contained within 90\% CI. In the case of $T_{\rm obs}=10\mathrm{yr}$, the system chirps, so the information provided by the likelihood dominates over the prior, therefore, this bias is corrected
and most of the degeneracies (at least partially) broken. In general, the posterior for the spins for weakly chirping systems are badly constrained and closely resemble the priors.
For chirping systems, the determination of spins can be understood from Fig.~\ref{chischia}. Because of the orientation of lines $\chi_{\rm PN}=\rm{const}$, $\chi_1$ is better constrained than $\chi_2$. As the mass ratio increases the slope of these lines changes, accentuating this difference. Spins of same (opposite) sign, are better (worse) determined as their magnitude increase because of the narrowing (broadening) of the allowed region. For the \emph{Fiducial} system, the error on the spin of the primary BH is quite large but we can infer that the spin is positive with $0$ (and negative values) being outside the $90 \%$ CI given in Table \ref{errors}.
The effective spin is measured within $0.1$ for chirping systems.

%As discussed above, when we observe the system at higher frequencies $\chi_{\rm PN}$ is well determined and the different spin configurations do not impact the masses measurement. When observing at lower frequencies only, $\chi_{\rm PN}$ is poorly constrained, and due to its strong correlation with $\eta$ it can affect the mass measurement.
%This is the case for the \emph{Spinup} system in the $T_{\rm obs}=4 \  years$ case as we show on the left panel of Fig. \ref{degen_chis_eta}. The prior pushes the posterior distribution towards $\eta=0.25$ and $\chi_+=0$ whereas the likelihood pushes towards the true value of $\chi_+$ (0.95). We end up with a ``compromise'', the $\chi_+$ distribution peaks between 0 and 0.95 and because of how its correlation with $\chi_+$ is oriented, $\eta$ is pushed towards lower values, peaking away from the injected value. As a consequence, the injected masses are almost outside the support of the posterior as seen on the right panel of Fig. \ref{degen_chis_eta}. However, with our definitions they are in the $90 \%$ CI. In the $T_{\rm obs}=10 \  years$ case,  we observe the chirp of the system, so the information provided by the likelihood dominates over the prior and this bias is corrected.

\
%=======
%As discussed above, when we observe the system at higher frequencies $\chi_{\rm PN}$ is well determined and the different spin configurations do not impact the masses measurement. When observing at lower frequencies only, $\chi_{\rm PN}$ is poorly constrained, and due to its strong correlation with $\eta$ it can affect the mass measurement.
%This is the case for the \emph{SpinUp} system in the $T_{\rm obs}=4 \  years$ case as we show on the left panel of Fig. \ref{degen_chis_eta}. The prior pushes the posterior distribution towards $\eta=0.25$ and $\chi_+=0$ whereas the likelihood pushes towards the true value of $\chi_+$ (0.95). We end up with a ``compromise'', the $\chi_+$ distribution peaks between 0 and 0.95 and because of how its correlation with $\chi_+$ is oriented, $\eta$ is pushed towards lower values, peaking away from the injected value. As a consequence, the injected masses are almost outside the support of the posterior as seen on the right panel of Fig. \ref{degen_chis_eta}. However, with our definitions they are in the $90 \%$ CI. In the $T_{\rm obs}=10 \  years$ case,  we observe the chirp of the system, so the information provided by the likelihood dominates over the prior and this bias is corrected.
%>>>>>>> 3b13580600606687bafd3cb0c5aad7c332dbb382

All results (for masses) so far were for redshifted masses. Since $\mathcal{M}_{c,s}=\chirpm/(1+z)$, we get:

\begin{equation}
 \frac{\Delta \mathcal{M}_{c,s}}{\mathcal{M}_{c,s}}= \frac{\Delta z}{1+z} + \frac{\Delta \chirpm}{\chirpm}. \label{rel_source_mass}
\end{equation}

As we discuss in Sec.~\ref{ext2}, $D_L$ is typically measured within $40$--$60 \%$ which implies a measurement of the redshift $z$ within $\sim 40$--$60 \%$ (at the low redshifts we are considering, $D_L$ and $z$ are linearly related). Thus, the second term on the right-hand side of Eq.~\eqref{rel_source_mass} is clearly subdominant and the error on the source frame chirp mass is dominated by the error on redshift, as a result we get:
\begin{equation}
 \frac{\Delta \mathcal{M}_{c,s}}{\mathcal{M}_{c,s}} \simeq \frac{0.5 z}{1+z}.
\end{equation}
This error is typically of the order of a few percent for systems detectable by LISA (up to $z \sim \mathcal{O}(10^{-1})$), which is better than current LIGO/Virgo measurements \cite{LIGOScientific:2018mvr}. This estimate is in good agreement with the results presented in Table \ref{errors}. The error for individual masses in the source frame are dominated by the error on the masses (like the second term on the left-hand side of Eq.~\eqref{rel_source_mass}) due to poorly constrained mass ratio.

%In Fig. we show the contribution of the different PN orders for $f \in [f_0,f_{max,LISA}]$ and zoom on the range $[f_0,f_{4 \ years}]$.

The initial frequency is always extremely well determined, with relative errors below $10^{-5}$. Its determination improves for chirping systems due to reduction of correlation with other intrinsic parameters. The frequency of the system at the beginning of the LISA observations is coincidental, as it is directly linked to the time that is left for the system to coalesce. We apply the stationary phase approximation~\eqref{eq:tf} to the full GW phase to infer $t_c$. This transformation involves all intrinsic parameters, so the error on $t_c$ is typically smaller for chirping systems. We find an error of the order of 1 day for systems far from merger, while for more strongly chirping systems $t_c$ can be determined to within $30 \ {\rm s}$.

We find that increasing or decreasing the total mass of the system (while preserving the SNR) as in the \emph{Heavy} and \emph{Light} systems has little consequence on the estimation of intrinsic parameters. The error on spins and symmetric mass ratio are the same as in the \emph{Fiducial} case. The relative error on chirp mass and initial frequency is slightly smaller for lighter systems (factor $\simeq 1.4$ between the \emph{Heavy} and \emph{Light} systems) because of the larger number of cycles. However, we do not find a simple scaling with the chirp mass of the system for a fixed level of SNR. In particular, we do not find the error on the chirp mass to scale with $\chirpm^{5/3}$ as computed in \cite{Finn:1992xs,Cutler:1994ys}. This was to be expected since as discussed in this section the error on intrinsic parameters depends crucially on the frequency interval through which we observe the binary.
%Remember that we rescaled the distance so that the SNR for a mission duration of 10 years is the same than for the \emph{Fiducial} system.

Finally, the choice of prior only marginally affects the posterior distribution for chirping systems. On the other hand, it can has a significant impact for nonchirping systems as can be seen in Fig.~\ref{priors_post}. For example,
the \emph{Flatmag} prior completely dominates the posterior distribution of spins as the KL divergences suggested
and shown in Fig.~\ref{priors_post}. Because of the noted correlations, the prior on spins propagates into the determination
of mass ratio and individual masses.
% (they are almost identical to Fig. \ref{comp_priors}) and the strong peaking at $\chi_{+,-}=0$ leads to a secondary peak at lower values of $\eta$. The \emph{Flatsampl} also favours null spins and unlike the other two priors, it does not favour equal mass ratios, thus because of the correlation with $\chi_+$ the distribution of $\eta$ is pushed towards lower values. In both cases the true value of $m_2$ is in the tail of the distrbution but with our definitions it is within the $90 \%$ CI. However, the true values of $\chi_+$ and $\chi_1$ are outside it when using the \emph{Flatmag} prior.

\subsection{Extrinsic parameters}

\begin{figure}
\centering
 \includegraphics[width=0.5\textwidth]{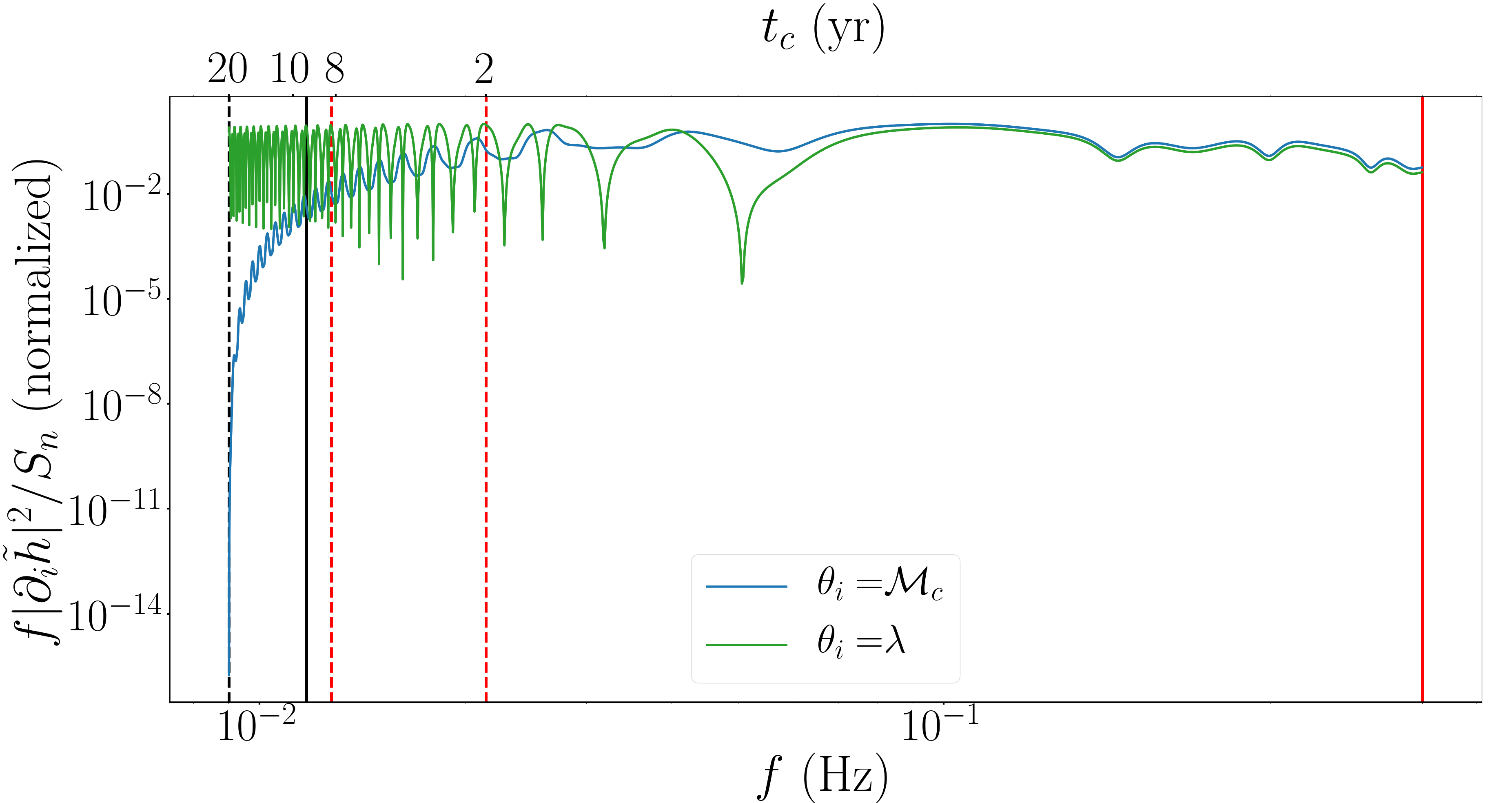}\\
 \centering
\caption{We plot $f|\partial_i \tilde{h}|^2/S_n$ (normalized to its maximum value) as a function of the time to coalescence. As discussed in Sec.~\ref{res_int}, most of the information on intrinsic parameters comes from the high end of the frequency band, whereas the contribution to sky parameters mainly comes from the low frequencies.}\label{dampl}
\end{figure}

\subsubsection{Sky location}\label{skyloc}

The sky location of the source is very well determined and, except for systems close to the equator, its posterior distribution is very similar to a Gaussian unimodal distribution.
We define the solid angle as in \cite{Cutler:1997ta}:
\begin{equation}
 \Delta \Omega=2 \pi \sqrt{(\Sigma^{\lambda,\lambda}) (\Sigma^{\sin(\beta),\sin(\beta)})-(\Sigma^{\lambda,\sin(\beta)})^2}\,, \label{eq_omega}
\end{equation}
where $\Sigma$ is the covariance matrix. This defines a $63\%$ confidence region around the true location. The error for the \emph{Fiducial} system, reported in Table \ref{errors}, is below $0.4 \ {\rm deg^2}$ which is within the field of view of most planned electromagnetic instruments such as Athena and SKA. \citep{WinNT,2018CoSka..48..498M}. With the exception of the \emph{Equatorial} system, the sky position is constrained with a similar precision for all systems considered in this work. The good localization comes from the complicated modulations imprinted on the signal by the orbital motion of LISA, according to~\eqref{kernel}. To understand how the sky localization evolves as a function of the frequency band we observe a system, in Fig.~\ref{dampl} we plot $f |\partial_\lambda \tilde{h}|^2/S_n$ (normalized with respect to its maximum value) as a function of the frequency. The quantity $f |\partial_i \tilde{h}|^2/S_n$ is the integrand entering the computation of the diagonal elements of the Fisher matrix~\eqref{def_fisher} and indicates (for each parameter) the most informative frequency range. Using a logarithmic scale for frequencies, the factor $f$ ensures that we can visualize the contributions to the integral as the area under the curve (up to a normalization factor). We also indicate the corresponding values of the time to coalescence $t_c$ on the upper x axis. We indicate the initial (dashed line) and end (solid line) frequencies of the \emph{Later} (red), \emph{Fiducial} (red) and \emph{Early} (black) systems for $T_{\rm obs}=10\mathrm{yr}$. The behavior for $\sin(\beta)$ is similar to $\lambda$. For comparison we show the same quantity for the chirp mass, the behavior for other intrinsic parameters is similar. As discussed in the previous section, most of the information on intrinsic parameters comes from high frequencies. On the other hand, there is more information on the sky location at low frequencies, where a given range of frequencies corresponds to more orbital cycles of the LISA constellation. However, this is to be balanced with the narrower frequency range spanned by systems evolving at lower frequencies, for a fixed observation time. For this reason, the \emph{Later} system gives a better localization than the \emph{Earlier} system even in the $T_{\rm obs}=10\mathrm{yr}$ case as reported in Table \ref{comp_err_sky} (0.05 against 0.2 ${\rm deg}^2$).

We can distinguish two main effects in~\eqref{kernel} informing us about the sky localization: the time dependency (through $t_{f}$, see Eq.~\eqref{eq:tf}) of the response reflects the orbital cycles of LISA, and the Doppler modulation $\exp(2i\pi f {\bf k} \cdot {\bf p}_0)$ of the phase. The Doppler modulation shows this time dependency, but also scales with $f$, so this term is larger for chirping signals reaching high frequencies. 
We find a better sky localization for lighter systems: $\Delta\Omega_{\emph{Light}}<\Delta\Omega_{\emph{Fiducial}}<\Delta\Omega_{\emph{Heavy}}$ (ranging from $0.1$ to $0.3$ ${\rm deg}^2$ in the $T_{\rm obs}=4\mathrm{yr}$ case and from $0.02$ to $0.05$ ${\rm deg}^2$ in the $T_{\rm obs}=10\mathrm{yr}$ case).
This is a result of keeping fixed the time to coalescence $t_c$ and the SNR (by adjusting the distance) for those systems. The GW signal from the lighter and heavier systems is displaced at higher and lower frequencies, since the evolution rate of the inspiral depends primarily on the chirp mass. Namely, $f_{0}=9.9 \, 12.7 \, 16.4 \ \mathrm{mHz}$ for \emph{Heavy}, \emph{Fiducial}, and \emph{Light}, respectively. Since we keep the SNR fixed in this comparison, this means that the lighter system has a stronger sky-dependent Doppler modulation of the phase, helping with the localization.

%with a factor $\sim 1.5$ between each of these terms, like the inverse ratio of their respective chirp mass.
%Although at first this might seem in contradiction with our previous discussion, it is not: most of the information on the source location comes from early frequencies (where the system spends more time), because we fixed the time to coalescence to 8 years these early frequencies are higher for lighter systems and terms in the response such as the Doppler phase are larger at higher frequencies, improving the discriminative power. In other words: because we rescaled the distance to keep a fixed level of SNR, the quantity $|\partial_\lambda(\tilde{h})|/f S_n$ is larger enough for lighter systems to compensate for the slightly narrower frequency interva spanned by lighter systems.

\begin{figure}
\centering
 \includegraphics[width=0.5\textwidth]{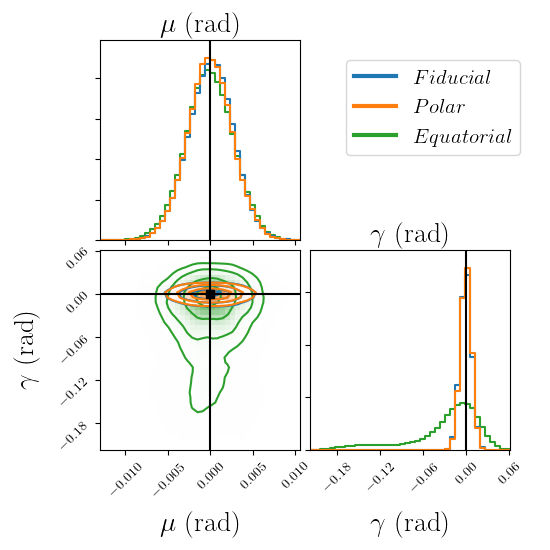}\\
 \centering
 \caption{Inferred distribution on the angles parametrizing the position of the source for the \emph{Polar}, \emph{Fiducial} and \emph{Equatorial} systems, with $T_{\rm obs} = 4\mathrm{yr}$. As explained in the main text, to avoid coordinate effects near the pole we do not compare the angles in the SSB frame ($\lambda,\beta$) but transformed angles ($\mu,\gamma$) defined by placing the injection point at the equator in each case (note that the scale of the two axis is not the same). The injection corresponds to $\mu=\gamma=0$ as indicated by the black solid lines and squares. $\mu$ is equally well recovered in the three cases. For the \emph{Equatorial} system we find a tail extending to the position $\beta \to -\beta$.}\label{comp_beta}
\end{figure}

\begin{table}
  \begin{center}
   \begin{tabular}{c|c|c|}

   \cline{2-3}

   & \multicolumn{2}{|c|}{$\Delta \Omega \ {\rm (deg^2)}$}  \\

    \cline{2-3}

    & $T_{\rm obs}=4\mathrm{yr}$ & $T_{\rm obs}=10\mathrm{yr}$ \\

    \hline

    \multicolumn{1}{|c|}{\emph{Fiducial}} & $0.18$ & $0.03$ \\

    \hline

    \multicolumn{1}{|c|}{\emph{Earlier}} & $0.70$ & $0.20$ \\
    
    \hline

    \multicolumn{1}{|c|}{\emph{Later}} & $0.05$ & / \\
    
    \hline

    \multicolumn{1}{|c|}{\emph{Polar}} & $0.14$ & $0.02$ \\

   \hline

   \multicolumn{1}{|c|}{\emph{Equatorial}} & $2.74$ & $0.24$ \\

   \hline

\end{tabular}
   \end{center}
    \caption{Solid angle around the injection point corresponding to a $63\%$ confidence region, computed with~\eqref{eq_omega}. The sky localization is slightly better for the \emph{Polar} sky position $\beta \simeq \pm \pi/2$, but much worse for the \emph{Equatorial} sky position $\beta \simeq 0$. The sky localization is better for the \emph{Later} system than for the \emph{Earlier} system (despite a lower SNR in the $T_{\rm obs}=10\mathrm{yr}$ case) due to the broader frequency range spanned during its observation.}\label{comp_err_sky}
  \end{table}

When comparing the \emph{Polar}, \emph{Fiducial} and \emph{Equatorial} systems, a direct comparison of the sky localization could be quite misleading because the metric on a sphere depends on the latitude, with a singularity at the pole. To evade this issue we define a system of coordinates on the sphere ($\mu,\gamma$) such that the injection point is always on the equator. The transformation from the ecliptic coordinates to this frame is source dependent. The spherical coordinates at the equator are locally
Cartesian and simplify the comparison of the results. We show the results of the sky localization in Fig.~\ref{comp_beta} for the \emph{Polar}, \emph{Equatorial} and \emph{Fiducial} systems in the ($\mu,\gamma$) frame and for $T_{\rm obs}=4 \mathrm{yr}$. All three systems recover $\mu$ (azimuthal angle) similarly well but the determination of $\gamma$ worsens as $\beta \to 0$. Furthermore, for the \emph{Equatorial} system we find a tail extending all the way to a secondary sky position corresponding to $\beta \to -\beta$. This behavior is due to the dominant Doppler phase in the frequency response which goes as $\cos(\beta)$: although the amplitude of the effect itself is maximized, its variation with the latitude is minimal as $\cos(\beta)$ is flat for $\beta=0$. For $T_{\rm obs}=10\mathrm{yr}$ this partial degeneracy is broken thanks to a combination of effects: there are more cycles of LISA's orbit contributing, the signal reaches high frequencies where the $f$ dependent terms in the response~\eqref{kernel} are larger, and the total SNR itself is larger.
%and  additional modulations coming from terms other than the Doppler phase (like the ${\bf n}_l \cdot {\bf P}_{22} \cdot {\bf n}_l$ terms in Eq. \eqref{kernel}) which depend on $\beta$ differently.
% It becomes a stronger function of $\beta$ as $\beta$ departs from 0. Moreover, the reflection with respect to the equatorial plane is an exact symmetry for this term. In the $T_{\rm obs}=10 \ years$ case, the \emph{Equatorial} system evolves up to higher frequencies and other sky location dependent terms in the response give a relevant contribution, allowing to break this partial symmetry and to reduce the error on $\beta$.
The solid angle for the \emph{Equatorial} system is larger as compared to other systems (as reported in Table \ref{comp_err_sky}) but remains much tighter than the current sky localization with ground-based observatories \cite{LIGOScientific:2018mvr}.

  \begin{figure}
\centering
 \includegraphics[width=0.5\textwidth]{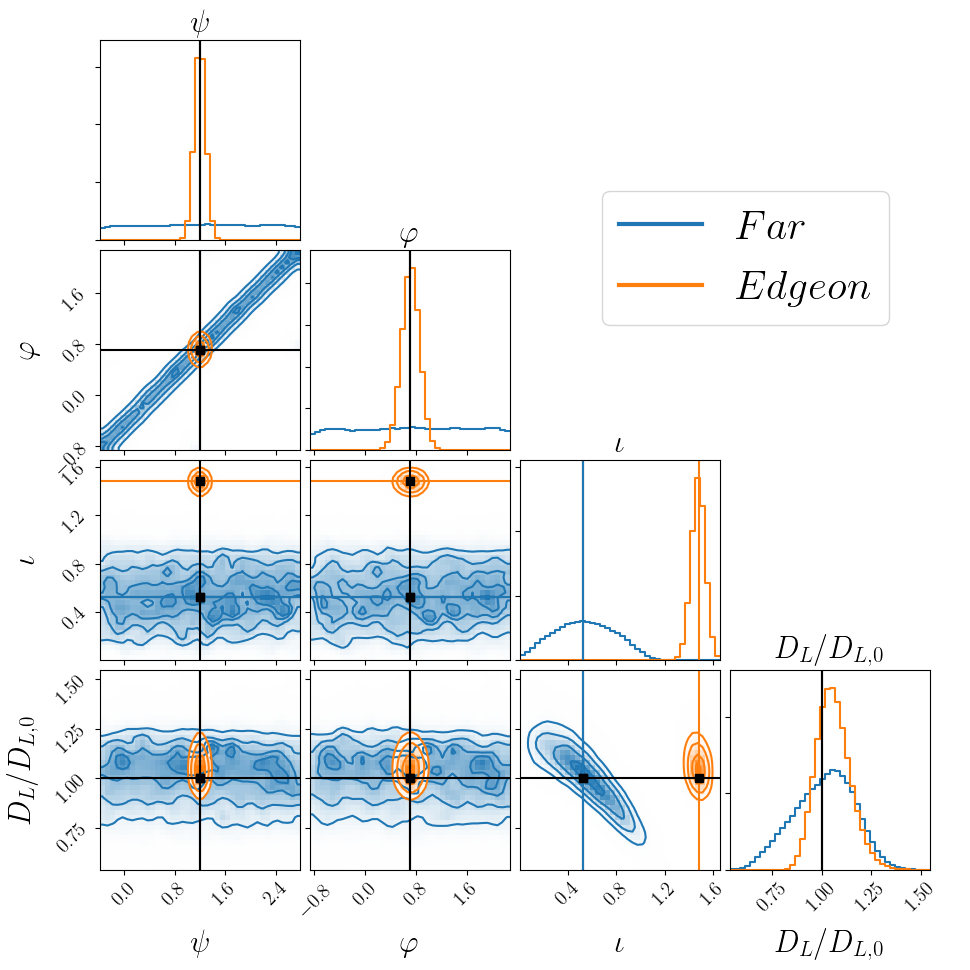}\\
 \centering
 \caption{Comparison of the inferred distribution on $\psi$, $\phi$, $\iota$ and $D_L$ for the \emph{Far} and \emph{Edgeon} systems, both have similar SNR. We normalized the distance to the injection value. Black lines and squares indicate the true values common for both systems and colored lines and squares the value of $\iota$ for each system. For the \emph{Edgeon} system the degeneracies between $\phi$ and $\psi$ and between $\iota$ and $D_L$ are broken giving a better estimation of each of these parameters. However, close to edge-on systems will usually have much lower SNR. Indeed, in order to keep a comparable SNR the distance to the \emph{Edgeon} is less than half the distance to the \emph{Far} system. %Thus, LISA will mostly observe face-on or face-off systems.
 }\label{comp_edgeon}
\end{figure}

%\subsubsection{$\psi,\varphi,\iota,D_L$}\label{ext2}
\subsubsection{Other extrinsic parameters}
\label{ext2}

We find strong correlations between inclination and distance, and between the polarization and the initial phase. These degeneracies are commonly seen in the analysis of LIGO/Virgo sources when using only the dominant $2,\pm 2$ mode in the analysis. With only the dominant $2,\pm 2$ mode, the GW in the radiation frame is given as:
%
% In the following $W$ stands for the wave as defined in \cite{Marsat:2020rtl}.
%Decomposing $h_{+,W}$ and $h_{\times,W}$ in spin weighted spherical harmonics \cite{doi:10.1063/1.1705135} and keeping only the 22-mode gives:
%
%\begin{align}
% \tilde{h}_{+,W}(f)&= \frac{1}{2}\left( \right . (_{-2}Y_{22}+_{-2}Y_{2-2}^{*}) \tilde{h}_{22,W} + \nonumber \\
%  &(_{-2}Y_{22}^{*}+_{-2}Y_{2-2}) \tilde{h}_{2-2,W} \left . \right ) \\
% \tilde{h}_{\times,W}(f)&= \frac{1}{2} \left (  \right . (_{-2}Y_{22}-_{-2}Y_{2-2}^{*}) \tilde{h}_{2-2,W} + \nonumber \\
% &(_{-2}Y_{22}^{*}-_{-2}Y_{2-2})  \tilde{h}_{2-2,W}^{*}  \left . \right ),
%\end{align}
%
%
%The exact expression for $_{-2}Y_{2\pm2}$ is \cite{Blanchet:2002av}:
%
%\begin{align}
% _{-2}Y_{22}=&\sqrt{\frac{5}{64 \pi}}(1+\cos \iota)^2e^{-2i\varphi} \\
% _{-2}Y_{2-2}=&\sqrt{\frac{5}{64 \pi}}(1-\cos \iota)^2e^{2i\varphi}.
%\end{align}
%
%Observe the minus sign for $\varphi$ relative to the usual definition due to our conventions.
%For non precessing binaries: $\tilde{h}_{2-2}=\tilde{h}_{22}^{*}$ so:
%
%\begin{align}
% \tilde{h}_{+,W}&=\sqrt{\frac{5}{4 \pi}}\frac{(1+\cos^{2})}{2}{\mathcal Re}(\tilde{h}_{22,W}e^{-2i\varphi}) \label{eq_hp} \\
%  \tilde{h}_{+,W}&=\sqrt{\frac{5}{4 \pi}}\cos \iota{\mathcal Im}(\tilde{h}_{22,W}e^{-2i\varphi}). \label{eq_hc}
%\end{align}
%
%
%Defining $\tilde{h}_{22,W}=\sqrt{\frac{4\pi}{5}}A(f)e^{2i\Psi(f)}$, we get:
\bsub
\begin{align}
 \tilde{h}_{+} (f) &= \tilde{A}(f) \frac{1+\cos^{2}(\iota)}{2} e^{2 i \varphi} e^{-2 i \Psi(f)} \,,\\
  \tilde{h}_{\times} (f) & = i \tilde{A}(f) \cos (\iota) e^{2 i \varphi} e^{-2 i \Psi(f)} \,,
\end{align}
\esub
where $\tilde{h}_{22}(f) = A(f) \exp(-i\Psi (f))$ is the frequency domain amplitude and phase decomposition of the mode $h_{22}$, with $\tilde{A} \equiv \sqrt{5/16\pi} A(f) $ absorbing conventional factors. We refer to~\cite{Marsat:2020rtl} for notation; in particular we exploit the symmetry between $h_{22}$ and $h_{2,-2}$ for nonprecessing systems to write the waveform in terms of $h_{22}$ only. Going to the SSB frame we rotate by the polarization angle $\psi$:
\bsub\label{eq:hpcSSB}
\begin{align}
 \tilde{h}_{+}^{\rm SSB} &= \tilde{h}_{+}\cos(2\psi)-\tilde{h}_{\times}\sin(2\psi) \,,\\
 \tilde{h}_{\times}^{\rm SSB} &= \tilde{h}_{+}\sin(2\psi)+\tilde{h}_{\times}\cos(2\psi) \,.
\end{align}
\esub

For a face-on system, $\iota=0$ leading to:
\bsub
\begin{align}
	\tilde{h}_{+}^{\rm SSB}(f) &= \tilde{A}(f) e^{2 i (\varphi - \psi)} e^{-2 i \Psi(f)} \,, \\
	\tilde{h}_{\times}^{\rm SSB}(f) &= i\tilde{A}(f) e^{2 i (\varphi - \psi)} e^{-2 i \Psi(f)} \,,
\end{align}
\esub
Thus we see that the initial phase and the polarization appear only through the combination $\psi-\varphi$ yielding a true degeneracy corresponding to $\psi-\varphi=\rm{const}$. For systems close to face-on/face-off, like the \emph{Fiducial} system, this gives the strong correlation between $\psi$ and $\varphi$ well observed in Fig.~\ref{comp_tobs}.
%For close to face-off systems ($\iota=\pi$), we would have a strong correlation corresponding to  $\psi+\varphi=cst$.
For edge-on systems ($\iota=\pi/2$) we have instead:
\bsub
\begin{align}
	\tilde{h}_{+}^{\rm SSB}(f) &= \tilde{A}(f) \cos (2\psi) e^{2 i \varphi} e^{-2 i \Psi(f)} \,, \\
%  \nonumber \\
% &=\frac{A(f)}{2} \left [ \cos(2\Psi(f)+2(\psi-\varphi)) \nonumber \right .\\
% &+\cos(2\Psi(f)-2(\psi+\varphi)) \left . \right ] \\
	\tilde{h}_{\times}^{\rm SSB}(f) &= \tilde{A}(f) \sin (2\psi) e^{2 i \varphi} e^{-2 i \Psi(f)} \,,
%  \nonumber \\
% &= \frac{A(f)}{2} \left  [ \sin(2\Psi(f)+2(\psi-\varphi)) \right . \nonumber \\
% &-\sin(2\Psi(f)-2(\psi+\varphi)) \left . \right ].
\end{align}
\esub
and the degeneracy between $\psi$ and $\varphi$ is then broken as also shown in Fig.~\ref{comp_edgeon}.
%that in this case both combinations $\psi+\varphi$ and $\psi-\varphi$ appear
%and we can determine both individually
 There we compare the distributions of $\psi$, $\varphi$, $\iota$ and $D_L$ for the \emph{Edgeon} system to the \emph{Far} system (which is almost face-on). Those systems have similar SNR. When the degeneracy between $\psi$ and $\phi$ is broken, we observe a correlation between the initial phase and the initial frequency. This is an artificial correlation due to relating $\varphi$ to the value of the phase at $f_0$ for each template. Using a fixed reference frequency, such as the initial frequency of the injected signal for example, eliminates this correlation.

\begin{figure}
	\centering
	\includegraphics[width=0.5\textwidth]{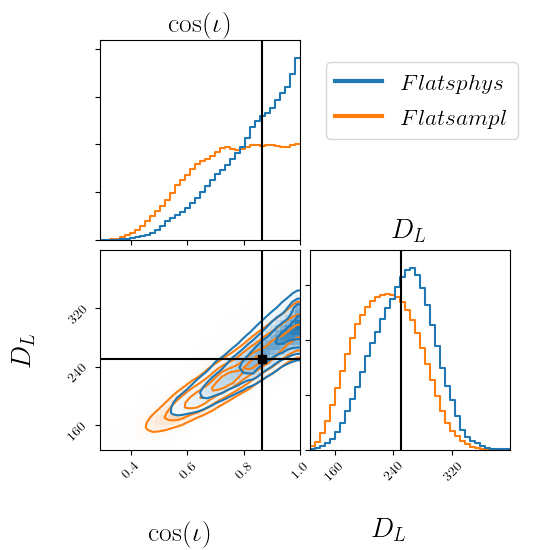}\\
	\centering
	\caption{Distribution of $\cos (\iota)$ and $D_L$ using the \emph{Flatphys} and \emph{Flatsampl} priors for $T_{\rm obs}=4\mathrm{yr}$. Although the distributions look rather different, the width of the $90\%$ CI for $D_L$ is barely affected and the true point, indicated by black lines and squares, is well within the CI.}\label{prior_dl}
\end{figure}

In Fig.~\ref{comp_edgeon} we also plot distance and inclination, which show a significant correlation for the
\emph{Far} system that is absent for the \emph{Edgeon} system. Distance and inclination are purely extrinsic parameters, and the degeneracy features when subdominant (higher order) modes are negligible appear in the same way for LIGO/Virgo and LISA. For LISA, see e.g., a discussion in the context of galactic binaries in~\cite{Shah:2012vc}. In short, in the limit of face-on/off systems the inclination acts as a scaling factor over a rather broad range of inclination values, thus changes in $\cos (\iota)$ can be compensated by changes in $D_L$. For close to edge-on systems, the $\times$ polarization of the wave is suppressed (in the wave-frame, before transforming to the SSB frame as in~\eqref{eq:hpcSSB}). The important point is that this suppression of $h_{\times}$ depends itself quite rapidly on the inclination, so that reproducing the injected signal leads to a rather tight constraint on $\iota$, and as a consequence on $D_L$. For MBHB observations with LISA, higher modes play an important role and help break these degeneracies \cite{Marsat:2020rtl}; but SBHBs are observed by LISA far from coalescence and higher modes are negligible for these signals.
%Both are the consequence of the absence of higher modes as we now explain.
%It is also apparent on this figure that the strong correlation between inclination and distance is partially broken for edge-on systems, decreasing the errors on both. The reason for that is well discussed in the context of galactic binaries in \cite{Shah:2012vc} and the argument holds here. In short when only the 22-mode is present, for face-on or face-off systems the inclination acts almost exclusively as a scaling factor, thus changes in $\cos \iota$ can be compensated by changes in $D_L$. For edge-on systems the inclination also affects the shape of the signal, leading to better constraint on $\iota$, and a consequence on $D_L$. We see from these equations that the waveform is a function of $\cos \iota$ rather than the inclination itself (this is what motivated our choice to use it as sampling parameter) thus we find it instructive to display the posterior distribution for $\cos \iota$ and how it is affected by the choice of prior.

\begin{figure*}
\centering
 \includegraphics[width=0.58\textwidth]{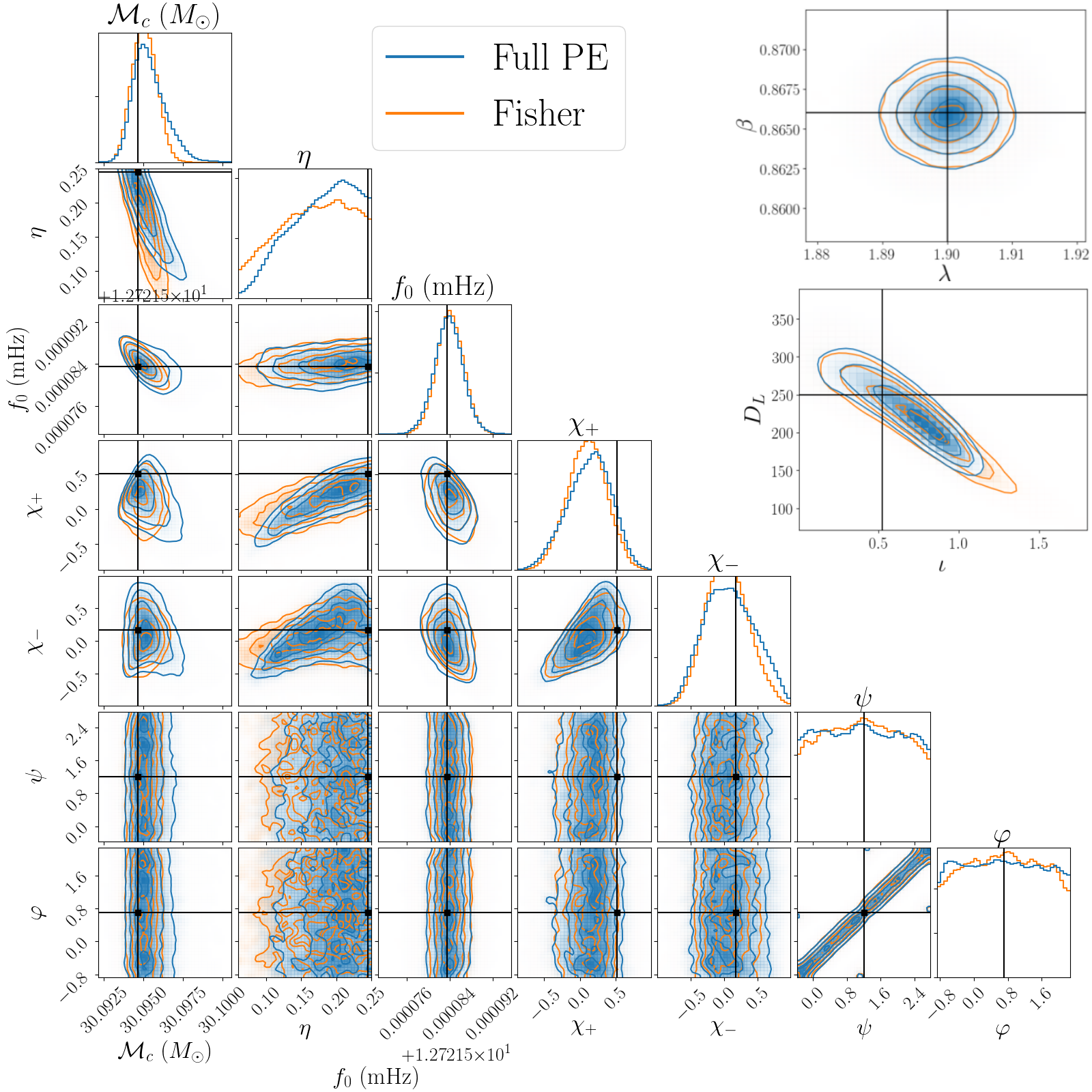}
 \centering
 \caption{Comparison between the inferred distribution for the \emph{Fiducial} system using the \emph{Flatsampl} prior and our Fisher analysis with $T_{\rm obs}=4\mathrm{yr}$. Black lines and squares indicate the true values.}\label{comp_fm4}
\end{figure*}

In Fig.~\ref{prior_dl} we show the effect of the distance prior on the posterior distribution for $\cos (\iota)$ and $D_L$ using the \emph{Flatphys} and \emph{Flatsampl} priors for $T_{\rm obs}=4\mathrm{yr}$. The former favors larger distances
 and, to keep the correct overall signal amplitude, compensates by preferring the face-on configuration. In the case of the \emph{Flatsampl} prior, the posterior distribution of $\cos (\iota)$ is flat because the likelihood itself is very flat around $\iota=0,\pi$ ($\cos (\iota)$ is a slowly varying function around its extrema).
% This is why Fisher-based-PE fails for $D_L$ and $\iota$ as we will show next and we have to truncate the eignevalues of the Fisher matrix to use it as a proposal.
Thus, the choice of prior shifts the peak of the posterior, but the $90\%$ CI still contains the true value and its width is largely unaffected.

Among all the cases we have considered, $D_L$ can be at best determined within $40 \%$, with the exception of the \emph{Edgeon} system for which we can determine distance to within $20 \%$. However, the edge-on systems will have lower SNR for a fixed distance to the source, and, therefore there is an observational selection effect where the face-on/off systems are preferred (that is what we observe with LIGO/Virgo). If we fix all other parameters of the \emph{Fiducial} system and set $\iota=\pi/2-\pi/36$, the SNR drops from $21$ to $9$ for $T_{\rm obs}=10\mathrm{yr}$.
%  We will only detect such systems if they are nearby ($D_L \simeq 150 \ {\rm Mpc}$) so most detected systems should be face-on or face-off. We find no drastic dependence on observing the system at higher frequencies as we did for intrinsic parameters and the sky location.
For the fixed inclination, time to coalescence and source position, the error on intrinsic parameters, distance and sky position scale, in first approximation, as $1/\mathrm{SNR}$.

\subsection{Fisher matrix analysis}\label{results_fm}

\begin{figure*}
\centering
  \includegraphics[width=0.58\textwidth]{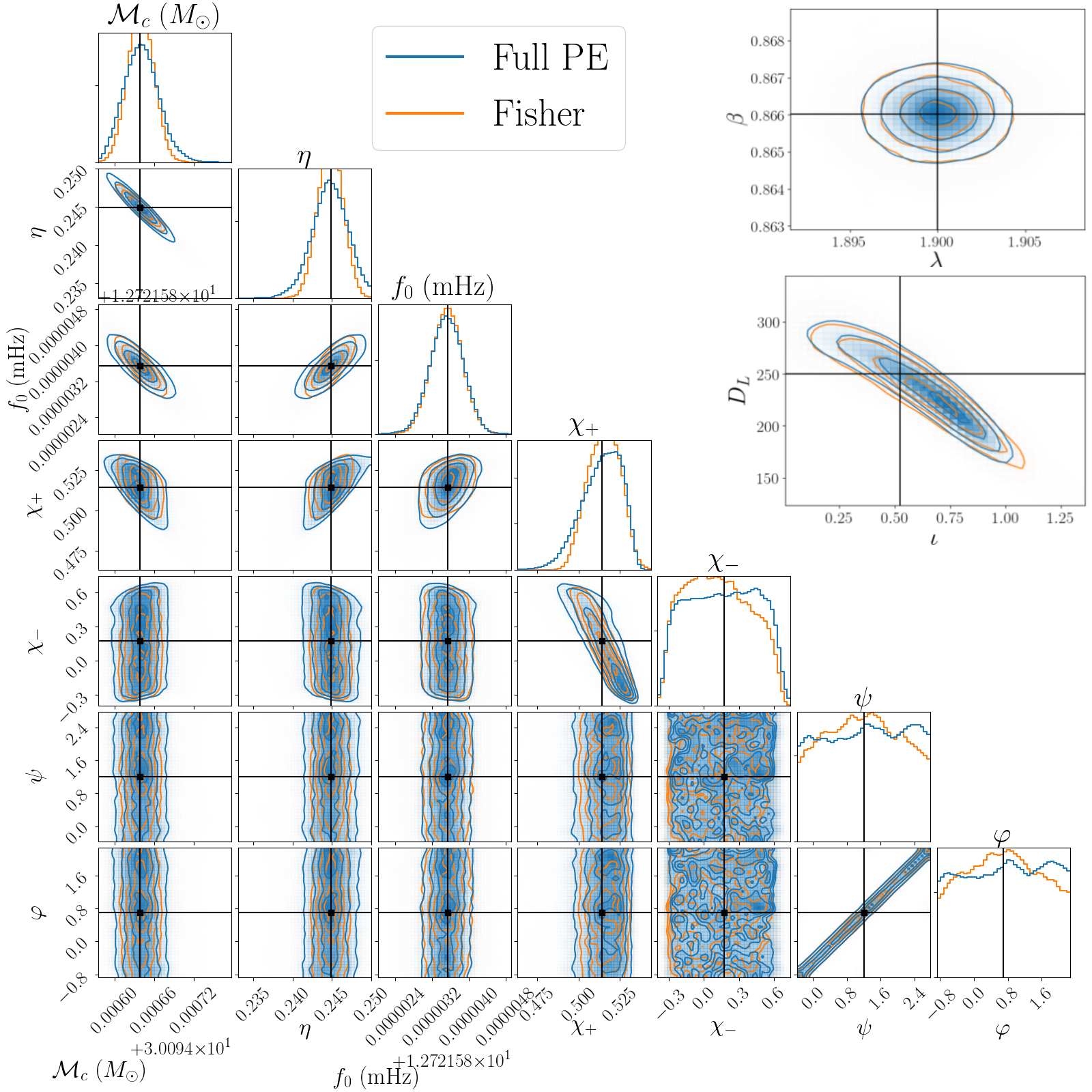}
 \centering
 \caption{Similar to Fig.~\ref{comp_fm4} but with $T_{\rm obs}=10\mathrm{yr}$. }\label{comp_fm10}
\end{figure*}

In this subsection we consider PE using a slightly improved version of Fisher information matrix analysis, inspired by \cite{Vallisneri:2007ev}. We have introduced the Fisher matrix in Sec.~\ref{fisher_mat} and discussed its augmented version, the effective Fisher, in Sec.~\ref{mhmcmc} for computing the covariance matrix. As we mentioned in Sec.~\ref{mhmcmc} and showed in Sec.~\ref{ext2}, the likelihood is very flat around $\iota=0,\pi$ leading Fisher-based-PE to overestimate the errors on $\cos (\iota)$ and $D_L$. To correct for this, we add an additional term ($F^{t}$) to the effective Fisher matrix: $F_{\rm eff}=F+F^{\rm p}+F^{\rm t}$ where $F$ is the ``original'' Fisher matrix given by Eq.~\eqref{def_fisher} and $F^{\rm p}$ is introduced to account for the prior on spins. Empirically, we found the choice $F^{\rm t}_{\cos (\iota),\cos (\iota)} = \frac{1}{(0.2(20/\mathrm{SNR}))^2}$ and 0 elsewhere to give good results for $\cos (\iota)$ and $\log_{10}(D_L)$. The prior matrix $F^{\rm p}$ does more than truncating the error on spins: it mimics the nontrivial prior on $\chi_{+}$ and $\chi_{-}$. Indeed, requiring the spins to be in the physically allowed range ($-1 \leq \chi_{1,2} \leq 1$) leads to a parabola-shaped prior on $\chi_{+,-}$ as seen on \ref{comp_priors}. We treat this nontrivial prior by a Gaussian distribution centered at $\chi_{+,-}=0$ with standard deviation $\sigma = 0.5$.
We invert the effective Fisher matrix to obtain the covariance matrix and use it to draw points from a multivariate Gaussian distribution. To fully account for the effect of the prior on spins, the point at which the Gaussian distribution is centered is shifted to $\theta_{\rm eff}=F_{\rm eff}^{-1}F\theta_{0}$. We only keep points within the boundaries given in Eq.~\eqref{flatsampl_prior}. For $\psi$ and $\phi$ we draw points in an interval of width $\pi$ around the central value.
In Figs. \ref{comp_fm4} and \ref{comp_fm10} we compare our Fisher analysis to the inferred distribution for the \emph{Fiducial} system using the \emph{Flatsampl} prior.

We find very good agreement despite the rather low SNR of this system, especially in the $T_{\rm obs}=4\mathrm{yr}$ case. In particular, the sky localization is the same for the full PE and the effective Fisher analysis. Naturally, this method cannot reproduce the secondary maximum we found for the \emph{Equatorial} system but it does predict a higher error as the system approaches the equatorial plane. The good agreement for $\chi_+$ and $\chi_-$ and for $T_{\rm obs}=4\mathrm{yr}$ is because the effective Fisher and posterior distribution are both prior dominated. For $\chirpm$ and $\eta$, Fisher agrees with the full PE on a 2-sigma level but cannot reproduce the bananalike correlation.
In case of $T_{\rm obs}=10\mathrm{yr}$, the likelihood becomes more informative for the effective spin, reducing the error predicted by the ``original'' Fisher while the $\chi_-$ distribution is still prior dominated.
Without adding $F^{\rm t}$ to the effective Fisher, the direction for the correlation between $\cos(\iota)$ and $D_L$ is predicted well but the Fisher matrix severely overestimates the error for nearly face-on/face-off systems. For the \emph{Edgeon} system, the likelihood is not so flat so the error predicted by the ``original'' Fisher is already small (in agreement with the Bayesian analysis) and adding $F^{\rm t}$ does not affect the PE.
 % flatness of the $\cos \iota$ distribution for nearly face-on system.
%The largest discrepancy between two approaches is in the estimation of the inclination and the distance. The direction
%for the correlation is predicted well but the Fisher matrix severely overestimates the error due to
 % flatness of the $\cos \iota$ distribution for nearly face-on system.
%
% close to the injection point is correct but due to the flatness of the $\cos \iota$ distribution, the error predicted by Fisher is way too large. In principle, we could apply the same trick we used for spins but it is not as clear what should be used as effective prior for $\iota$ and $D_L$.

\begin{figure*}
\centering
  \includegraphics[width=0.8\textwidth]{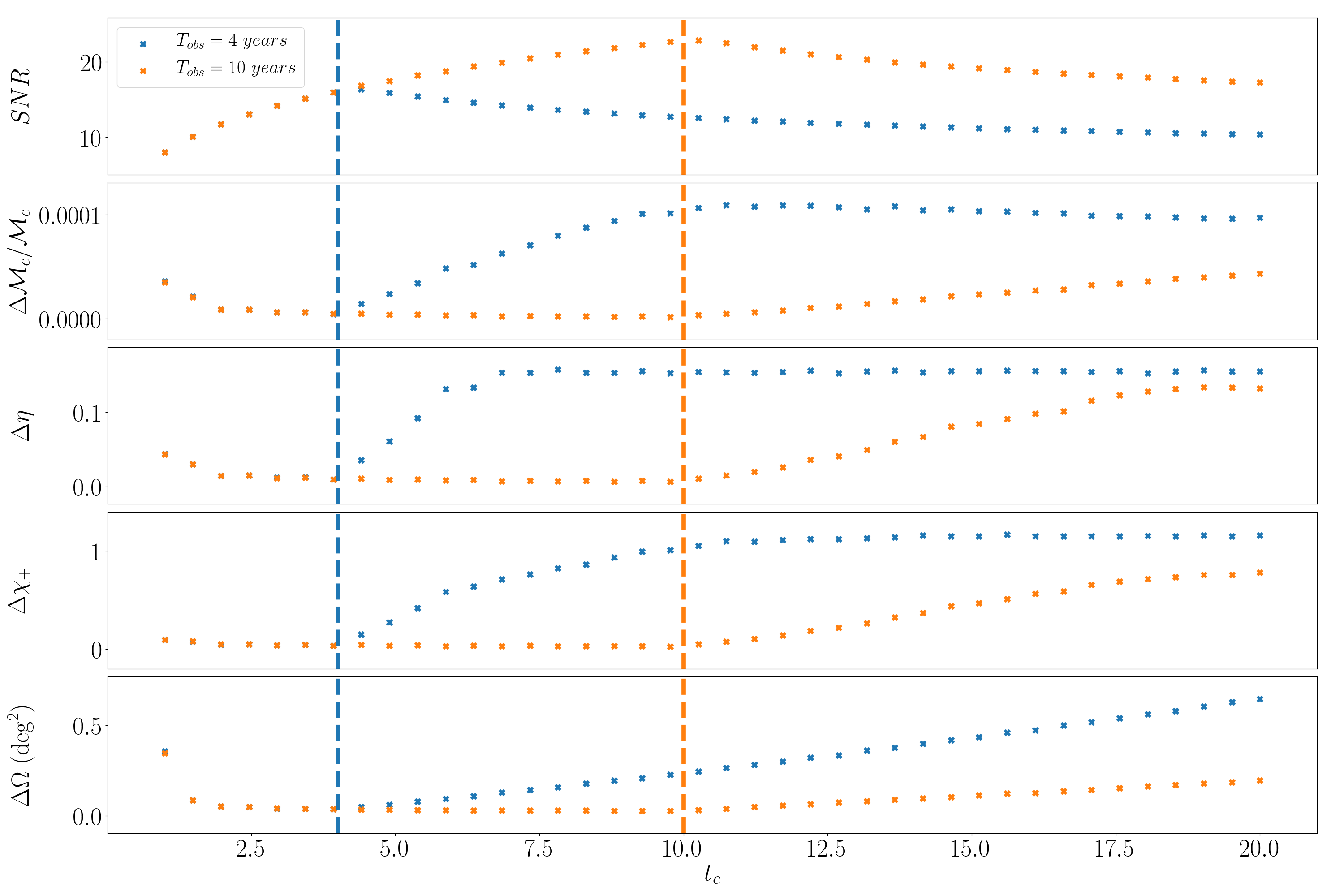}
 \centering
 \caption{Evolution of the error as function of the time before merger we start observing the system in the $T_{\rm obs}=4\mathrm{yr}$ and $T_{\rm obs}=10\mathrm{yr}$ cases. The SNR is given in the upper panel. The errors on $\chirpm$, $\eta$, and $\chi_+$ correspond to the width of the $90 \%$ CIs, and $\Delta \Omega$ is defined in \eqref{eq_omega}.}\label{err_fm}
\end{figure*}

Based on the rather good agreement we found with Bayesian PE, we can exploit the simplicity of Fisher analysis to further explore how does the PE evolve with the time (left) to coalescence. In Fig.~\ref{err_fm}, assuming $T_{\rm obs}=4\mathrm{yr}$ and $T_{\rm obs}=10\mathrm{yr}$, we plot the errors on $\chirpm$, $\eta$, $\chi_+$ and $\Delta \Omega$ as a function of the time to coalescence $t_c$, keeping all the parameters of the \emph{Fiducial} system fixed but varying the initial frequency in accordance with the chosen $t_c$. We plot the corresponding evolution of the SNR in the top panel,
with the lowest SNR of 8 being reached for $t_c\simeq1\mathrm{yr}$. Dashed lines mark $t_c=T_{\rm obs}$ in each case which corresponds to the maximum achievable SNR given the observation time and it also corresponds to the best estimation of parameters. Note two different regimes on the two sides of the dashed line: to the left, the PE is governed by the decrease in the signal duration in LISA band and reduction in SNR, while to the right the PE is determined mainly by the bandwidth of the signal spanned over the observation time. As discussed in Sec.~\ref{skyloc} the sky localization comes mainly from modulations caused by the motion of LISA, therefore it worsens rapidly if the system spends too little time in band (below 1 year).

% For all the parameters, as $t_c$ decreases the error decreases until $t_c=T_{\rm obs}$ which corresponds to the minimum of the errors for all the parameters. This is because we get to observe the later evolution of the system, which as we discussed contains most of the discriminative power. As $t_c$ decreases below $T_{\rm obs}$, the error increases back because the signal leaves the band before the end of the mission and we observe it less longer. However, the increase in error is slower to the left of $t_c=T_{\rm obs}$ than to right, because for $t_c \leq T_{\rm obs}$ although the signal in band becomes shorter and the SNR decreases we observe the chirp of the system. Because our ability to localise the systems depends on observing the system long enough, $\Delta \Omega$ increases rapidly as $t_c$  approaches one year but the sky location is till well measured.
%Unfortunately, because the results predicted by our Fisher analysis for $D_L$ are not reliable we cannot carry the same study for it.

\subsection{Long-wavelength approximation}

 \begin{figure*}
\centering
  \includegraphics[width=0.5\textwidth]{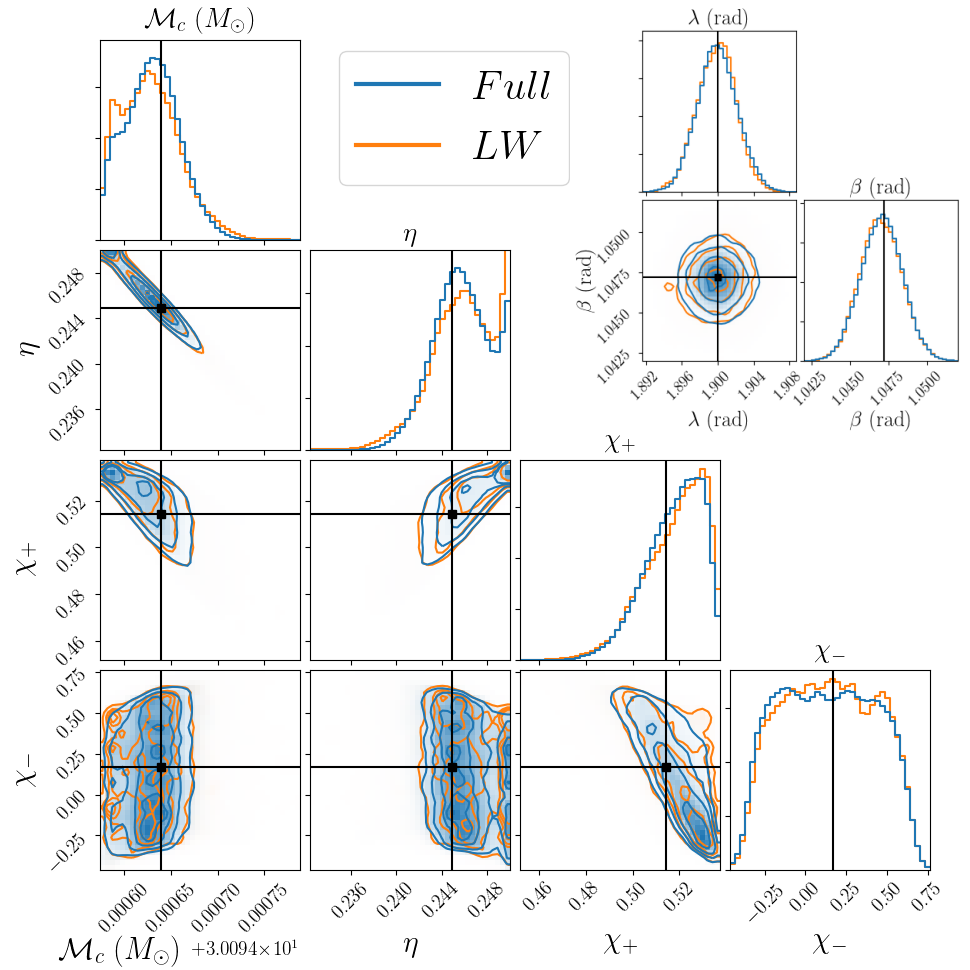}
 \centering
 \caption{Comparison of inferred distributions of intrinsic parameters and sky location using the \emph{Full} and \emph{LW} responses for the \emph{Fiducial} system in the $T_{\rm obs}=10\mathrm{yr}$ case. Black lines and squares indicate the true values.} \label{int_lw}
\end{figure*}

 %\begin{figure*}
%\centering
 %\includegraphics[width=0.3\textwidth]{comp_response_std_tobs4.png}
  %\includegraphics[width=0.3\textwidth]{comp_response_planar_tobs4.png}
   %\includegraphics[width=0.3\textwidth]{comp_response_polar_tobs4.png} \\
    %\includegraphics[width=0.3\textwidth]{comp_response_std_tobs10.png}
     %\includegraphics[width=0.3\textwidth]{comp_response_planar_tobs10.png}
      %\includegraphics[width=0.3\textwidth]{comp_response_polar_tobs10.png}
 %\centering
 %\caption{Comparison of inferred distributions of the sky location using the \emph{Full} and \emph{Lowf} responses for the \emph{Fiducial}. Top panel: $T_{\rm obs}=4\mathrm{yr}$ case, bottom panel: $T_{\rm obs}=10\mathrm{yr}$ case. Black lines and squares indicate the true values.} \label{sky_lw}
%\end{figure*}

\begin{figure*}
\centering
 \includegraphics[width=\textwidth]{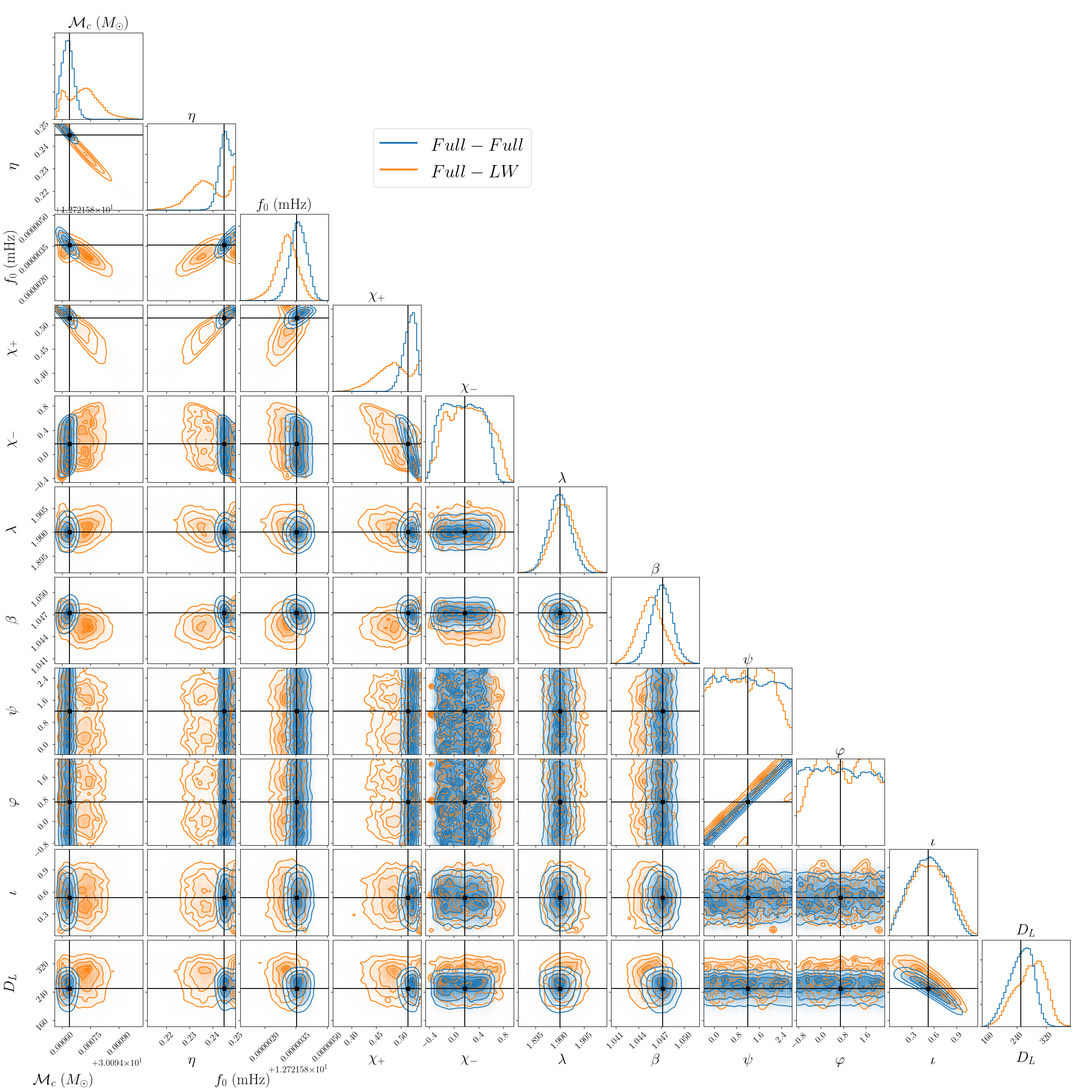}\\
 \centering
 \caption{Comparison of the inferred distributions for the \emph{Fiducial} system in the $T_{\rm obs}= 10\mathrm{yr}$ case using the \emph{Full} and the \emph{LW} response in the Bayesian analysis. In both cases, data was generated with the \emph{Full} response. Black lines and squares indicate the true values.}\label{comp_dt}
\end{figure*}

\begin{table}
  \begin{center}
   \begin{tabular}{c|c|c|c|c|}

   \cline{2-5}

  & \multicolumn{2}{|c|}{$T_{\rm obs}=4\mathrm{yr}$} & \multicolumn{2}{|c|}{$T_{\rm obs}=10\mathrm{yr}$} \\

    \cline{2-5}

    & \emph{Full} & \emph{LW} & \emph{Full} & \emph{LW} \\

    \hline

    \multicolumn{1}{|c|}{\emph{Fiducial}} & $13.5$ & $12.9$ & $21.1$  & $21.4$\\

    \hline

    \multicolumn{1}{|c|}{\emph{Polar}} & $12.8$ & $12.2$ & $20.1$ & $20.0$ \\

  \hline

  \multicolumn{1}{|c|}{\emph{Equatorial}} & $14.9$ & $14.2$  & $23.1$ & $23.4$\\

   \hline

   \end{tabular}
   \end{center}
    \caption{Comparison between the SNRs for the \emph{Fiducial}, \emph{Polar} and \emph{Equatorial} systems using the \emph{Full} and the \emph{LW} response.}\label{snrs_lw}
  \end{table}

 \begin{table*} [!ht]
  \begin{center}
   \begin{tabular}{c|c|c|c|c|c|c|}

   \cline{2-7}

   & \multicolumn{3}{|c|}{$T_{\rm obs}=4\mathrm{yr}$} & \multicolumn{3}{|c|}{$T_{\rm obs}=10\mathrm{yr}$} \\

     \cline{2-7}

    &  $\log \mathcal{L}(\theta_0)$ & max($\log \mathcal{L}$) & $\tilde{\rho}$ &  $\log \mathcal{L}(\theta_0)$ & max ($\log \mathcal{L}$) & $\tilde{\rho}$ \\

    \hline

   \multicolumn{1}{|c|}{\emph{Fiducial}}  & $-50$ & $-2$ & $0.99$ & $-268$ & $-38$ & $0.91$ \\

    \hline

   \multicolumn{1}{|c|}{\emph{Polar}}  & $-45$ & $-3$ & $0.99$ & $-234$ & $-30$ & $0.92$ \\

    \hline

   \multicolumn{1}{|c|}{\emph{Equatorial}}  & $-55$ & $-2$ & $0.99$ & $-288$ & $-34$ & $0.94$ \\

   \hline

\end{tabular}
   \end{center}
    \caption{Loglikelihood at the true point, maximum likelihood and relative SNR (defined in Eq.~\eqref{rel_snr}) when using the \emph{LW} approximation in the Bayesian analysis for data generated with the \emph{Full} response. }\label{snr_eff}
  \end{table*}

In Table \ref{snrs_lw} we compare the SNR for the \emph{Fiducial}, \emph{Polar} and \emph{Equatorial} systems using the \emph{Full} and \emph{LW} responses for two observation times. We find that accounting for the degradation at high frequencies (Eq.~\eqref{degrad_highf}), the \emph{LW} approximation seems to barely affect the PE as can be seen in Fig.~\ref{int_lw}. We find similar behavior for the \emph{Polar} and \emph{Equatorial} systems. Some care is needed in interpreting this result: this comparison shows that the high frequency terms neglected in the \emph{LW} approximation have little impact on the posterior of the sky position if the likelihood is computed self-consistently (signal and template are produced using same response, either \emph{LW} or \emph{Full}). However, when analyzing real data these high frequency terms cannot be neglected. In other words, these effects in the full response can indeed be subdominant in the parameter recovery, if more information comes from other effects like the LISA motion and the main Doppler modulation, while not being negligible in the signal itself. To illustrate this, we simulate data for the \emph{Fiducial}, \emph{Polar} and \emph{Equatorial} systems using the full response and perform a Bayesian analysis using the \emph{LW} approximation to compute templates. In Table \ref{snr_eff}, we give the log-likelihood evaluated at the true point, the maximum likelihood and the maximum overlap:
\begin{equation}
 \tilde{\rho}={\rm max}_h \left ( \frac{(d|h)}{\sqrt{(d|d)(h|h)}} \right ). \label{rel_snr}
\end{equation}
In practice, we compute the maximum overlap by optimizing over our samples. The quantity $1-\tilde{\rho}$ indicates how much SNR would be lost if wrong templates were used for the detection of signal. We find that up to $\sim 10\%$ of the SNR could be lost, given the already low SNR of SBHBs in LISA this would severely compromise our chances of detecting such sources. The very small value of the likelihood at the true point by itself shows that using the \emph{LW} approximation will have an impact on the PE. In Fig.~\ref{comp_dt}, we compare posterior distributions obtained by using template generated with \emph{Full} or \emph{LW} response while analyzing the \emph{Fiducial} system, with $T_{\rm obs}=10\mathrm{yr}$ and generated with the \emph{Full} response. This system has a significant bandwidth and the \emph{LW} template cannot fit simultaneously the low and high frequency content of the signal, causing severe biases in the PE and loss of SNR. The same system but with $T_{\rm obs}=4\mathrm{yr}$, shows different result, the \emph{LW} template is effectual enough to fit the signal rather well with the largest bias appearing only in $\psi-\varphi$ distribution as a compensation
for terms neglected in the response and with a mild drop in the SNR. However, those signals are quite weak and we do not have the luxury to loose even a small portion of SNR.

Thus, our findings seem to validate the \emph{LW} approximation for prospective PE studies, if it is used consistently for injecting and recovering the signal, while it would be inappropriate to analyze real data. However, we should remember that we did not explore the full parameter space, while \eqref{degrad_highf} is valid as an average over orientations, so a different choice of parameters could yield worse results. We also note that the full response~\eqref{kernel} is actually quite simple and not more expensive computationally, while being unambiguous and eliminating the need for the averaging entering \eqref{degrad_highf}.

% We find that for the signal can be well with an effectual template at the cost of shifting $\varphi$ (and $\psi$ as a consequence of the correlation between them) to compensate for the high frequency terms that are neglected in the response, yielding a similar $\psi-\varphi$ distribution to the one on Fig. \ref{comp_dt}. The high frequency terms neglected in the \emph{LW} approximation become more important and the dephasing between templates and data increases. Since a lot of the information on the source location comes from lower frequencies, it is less affected than one could imagine at first. On the other hand, most of the discriminative power on intrinisc parameters comes from high frequencies, therefore their estimation is signifificantly affected as can be seen on Fig. \ref{comp_dt}. The distance is also affected due to the increasing difference in amplitude between the \emph{Full} response and the \emph{LW} approximation at high frequencies.
%Thus, it makes no doubt that the full response has to be used when analysing real data.
%%Furthermore, since its computational cost is essentially the same than using the improved \emph{LW} approximation we presented in this work, we strongly recommend its use in future works.

\section{Discussion}\label{ccl}

Merging binary stellar mass BHs are detected almost weekly during the third LIGO/Virgo observational run (O3). In this work we explored what LISA will be able to tell us about those binaries. While ground-based detectors observe the last seconds before the merger, LISA will see the early inspiral evolution of those systems. The results of the O3 run are not publicly available yet, so we used a GW150914-like system as a fiducial system in our study. We varied the parameters of the system in turn, investigating the corresponding changes in PE. We constructed and analyzed simulated (noiseless) data applying the full LISA response. We employed a Bayesian PE analysis and cross-checked our results using two independent samplers.
We have found that PE results are most sensitive to the frequency span of the GW and its extent within LISA sensitivity region given the observation duration, or in other words, how much the signal chirps during the observation time.

For weakly chirping systems that do not reach high frequencies during LISA's observations, the GW phase is dominated by the leading PN order, with smaller contributions from higher PN terms. As a result, the best measured parameter is the chirp mass (entering at the leading order) with typical relative error below $10^{-4}$. The weak contributions of subleading terms up to 1.5 PN lead to a three-way correlation between spins, symmetric mass ratio and the chirp mass. The mass ratio is very poorly constrained and the posterior for the spins is dominated by the priors. We nonetheless recover the sky position very well (typically within $0.4$ deg$^2$) thanks to the amplitude and phase modulation of the GW signal due to LISA's motion. Such an area in the sky is within the field of view of electromagnetic instruments such as Athena and SKA.

For chirping systems that reach the high end of the LISA frequency band and coalesce during the observation, higher-order PN terms become more important and help break the correlations between intrinsic parameters, thus leading to a significant improvement in PE. The individual masses for chirping systems are measured within $20$--$30\%$ and even better for systems with higher mass ratio. The constraints on individual spins result from the combination of the measurement of the 1.5 PN spin combination and the physical boundaries of the prior on spins: $-1 \leq \chi_{1,2} \leq 1 $. This suggests using $\chi_{\rm PN}$ (specified in Eq.~\eqref{chipn}) as a sampling parameter. We find that the measurement of the time to coalescence improves as we observe the systems closer to merger, from $\mathcal{O}(1 \ {\rm day})$ (for mildly chirping binaries) to $\mathcal{O}(30 \ {\rm s})$. We note that the best way to increase our chances to observe SBHBs chirping is by increasing LISA's mission duration.

The measurement of the luminosity distance is less impacted by whether the systems are chirping, and it is essentially a function of their SNR. Much like LIGO/Virgo observations when higher modes can be neglected, the degeneracy between distance and inclination is important. In our example, the distance is typically measured within $40$--$60\%$ if the system is close to face-on/off and within $20 \%$ if the system is edge-on (when adjusting the distance to keep the same SNR). The distance and therefore redshift uncertainty dominates the measurement of the source-frame chirp mass at a percent level. The precision on individual masses in the source frame is dominated by intrinsic parameter degeneracies.

We have suggested an augmentation of the usual Fisher matrix approach, that we called the effective Fisher matrix, and we have shown that it gives rather reliable results for the sky position and intrinsic parameters of the system when compared to Bayesian PE. We also showed that combining the use of the long-wavelength approximation for LISA with the introduction of a degradation factor at high frequencies yields very similar results as compared to using the full response for computing likelihoods self-consistently (using the same response for injected data and templates). However, using the long-wavelength approximation to analyse real data could decrease the effective SNR by $10\%$, drastically reducing our chances of detecting the signals and has a significant impact on the PE, particularly on the measurement of intrinsic parameters. The computational cost of the full response being essentially the same as the long-wavelength approximation, we recommend its use in future work.

We can utilize the knowledge and understanding obtained in the study of PE for the development of search tools: (i) the PE for these systems is mainly unimodal, with secondary modes appearing either in special cases (like the \emph{Equatorial} system) or under the effect of priors where the likelihood is weakly informative (like the symmetric mass ratio for nonchirping systems); (ii) the chirp mass and the sky coordinates are the best measured parameters, so we can make a hierarchical search starting with those parameters and taking into account the correlations which are explored and understood in Sec.~\ref{results}; (iii) the effective Fisher is an efficient proposal for a Bayes-based search once we start to find indications for a candidate GW signal in the data. In addition, we can perform an incremental analysis starting with a half-year-long data segment and progressively increasing it. This works as a natural annealing scheme and should help in detecting (especially) chirping systems.

Detection of even few SBHBs by LISA which merge somewhat later with a very high SNR in the band of ground-based detectors \cite{Gerosa:2018wbw} will constitute ``golden events''. Beyond all the benefits of multiband detections {\it per se}, the information provided by LISA itself will be very valuable. For example, \cite{Tso:2018pdv} suggested that the measurement of the time to coalescence could be used to inform ground-based detectors and improve BH spectroscopy. The good estimate on the time to coalescence and the sky location could be used for electromagnetic follow-up of the source as suggested in \cite{Caputo:2020irr}. Finally, these measurements could be used to tighten the constraints on the Hubble constant ($H_0$) even if no electromagnetic counterparts are detected, using galaxy catalogues \cite{Schutz:1986gp,DelPozzo:2017kme,Kyutoku:2016zxn}. Moreover, we expect our results to extend to more massive systems such as ``light'' intermediate mass black holes binaries (IMBHB), i.e., with component masses $\mathcal{O}(10^2 M_{\odot}) $ and systems similar to the massive binary recently announced by the LIGO/Virgo collaboration \cite{Abbott:2020tfl,Toubiana:2020drf}.  

Modifications to the GW phase induced by either modified theories of gravity or environmental effects will generically involve additional coefficients parametrizing the underlying mechanisms and their correlation with the parameters of the system ($\chirpm$, $\eta$, \dots). As a consequence, even for low-frequency modifications, the best constraints or measurements will come from chirping systems, as we found in \cite{Toubiana:2020vtf} in the context of testing modified theories of gravity with LISA observations.

A major improvement to this work would be the inclusion of orbital eccentricity. Astrophysical formation models predict that binaries formed dynamically should have large eccentricities \cite{Antonini:2012ad,Samsing:2017xmd}. However, by the time these binaries reach the frequency band of ground-based detectors, they will have circularized. Thus, LISA could play an important role in the discrimination between different formation channels \cite{Samsing:2018isx,Nishizawa:2016jji,Nishizawa:2016eza,Breivik:2016ddj}. Furthermore, neglecting eccentricity could affect the PE and detection efficiency. We are currently limited by the lack of fast eccentric waveforms, but work is ongoing in this direction \cite{Cao:2017ndf,Ireland:2019tao,Hinder:2017sxy,Hinderer:2017jcs,Huerta:2017kez}.
Concerning spins, binaries formed dynamically are expected to have misaligned spins \cite{Gerosa:2018wbw}, causing the binary's orbit to precess. The system might endure a sizeable number of precession cycles over the lifetime of LISA, albeit with a small opening angle of the precession cone for a binary in its early inspiral. Precession effects can become more important close to merger and therefore should be considered carefully when relating the signal in the LISA band with the signal in the ground-based detectors band. We leave the investigation of precession effects for future work.

We conclude with the claim that this work, together with \cite{Toubiana:2020vtf, Caputo:2020irr}, has confirmed the scientific potential of the observation of SBHBs with LISA and should be seen as a first step towards an extensive study of the PE for multiband observations.

\section*{Acknowledgements}
A.T. is grateful to Nikolaos Karnesis for his help at the start of the project and his support all along it. We thank the anonymous referee for making useful suggestions. This work has been supported by the European Union's Horizon 2020 research and innovation program under the Marie Sk\l{}odowska-Curie Grant Agreement No 690904.
A.T. acknowledges financial support provided by Paris-Diderot University (now part of Université de Paris). The authors would also like to acknowledge networking support from the COST Action CA16104.

 \FloatBarrier

\bibliography{sobhb.bib}

\end{document}